\documentclass[12pt]{article}
\pdfoutput=1
\usepackage[nosort]{cite}

\usepackage{caption}

\usepackage{mwe}

\usepackage{draft} 
\usepackage{hyperref}
\usepackage{graphicx,color,subfig,upgreek,mathtools}
\usepackage{amsfonts,amssymb,amsmath,multirow}
\usepackage{mathabx,epsfig}
\usepackage{pifont}
\usepackage{cite}
\usepackage{mciteplus}
\usepackage{skak}
\DeclareFontFamily{OT1}{pzc}{}
\DeclareFontShape{OT1}{pzc}{m}{it}{<-> s * [1.10] pzcmi7t}{}
\DeclareMathAlphabet{\mathpzc}{OT1}{pzc}{m}{it}

\usepackage{bbold}

\def\be#1\ee{\begin{align}#1\end{align}}

\def\CA{{\mathcal A}}
\def\CB{{\mathcal B}}
\def\CC{{\mathcal C}}

\def\CG{{\mathcal G}}

\def\CL{{\mathcal L}}
\def\CM{{\mathcal M}}
\def\CN{{\mathcal N}}
\def\CO{{\mathcal O}}

\def\CT{{\mathcal T}}

\def\CZ{{\mathcal Z}}

\def\C{\mathbb{C}}

\def\Z{\mathbb{Z}}

\def\tilde{\widetilde}
\renewcommand{\bar}{\overline}

\def\^{{\wedge}}
\def\*{{\star}}

\definecolor{ao}{rgb}{0.13, 0.55, 0.13}

\newcommand{\Stab}{\text{Stab}}

\newcommand{\Orb}{\text{Orb}}

\newcommand{\zero}{{(0)}}
\newcommand{\one}{{(1)}}
\newcommand{\two}{{(2)}}

\newcommand{\reg}{\text{reg}}

\newcommand{\Trt}{\text{Tr}_t}

\newcommand{\tG}{\widetilde{G}}
\newcommand{\tT}{\widetilde{T}}
\newcommand{\half}{\frac{1}{2}}

\usepackage{tikz-cd}
\usepackage{datetime}
\begin{document}

\begin{titlepage}

\preprint{CALT-TH-2021-041 \\
UTTG 23-2021}

\begin{center}

\hfill \\
\hfill \\
\vskip 1cm

\title{Symmetries of 2d TQFTs and Equivariant Verlinde Formulae for General Groups}

\author{Sergei Gukov$^1$, Du Pei$^{2}$, Charles Reid$^3$, Ali Shehper$^{4,5}$
}

\address{$^{1}$Walter Burke Institute for Theoretical Physics, California Institute of Technology,
Pasadena, CA 91125, USA}

\address{$^2$Center of Mathematical Sciences and Applications, Harvard University, Cambridge, MA 02138, USA}

\address{$^3$Department of Mathematics, University of Texas, Austin, TX 78712, USA}

\address{$^4$Department of Physics, University of Texas, Austin, TX 78712, USA}

\address{$^5$Department of Mathematics, Yale University, New Haven, CT 06520, USA}


\end{center}

\abstract{We study (generalized) discrete symmetries of 2d semisimple TQFTs. These are 2d TQFTs whose ``fusion rules'' can be diagonalized. We show that, in this special basis, the 0-form symmetries always act as permutations while 1-form symmetries act by phases. This leads to an explicit description of the gauging of these symmetries. One application of our results is a generalization of the equivariant Verlinde formula to the case of general Lie groups. The generalized formula leads to many predictions for the geometry of Hitchin moduli spaces, which we explicitly check in several cases with low genus and $SO(3)$ gauge group. 
}

\end{titlepage}

\eject

\tableofcontents

\unitlength = .8mm

\setcounter{tocdepth}{3}

\section{Introduction}

\begin{figure}
\centering
        \includegraphics[totalheight=5cm]{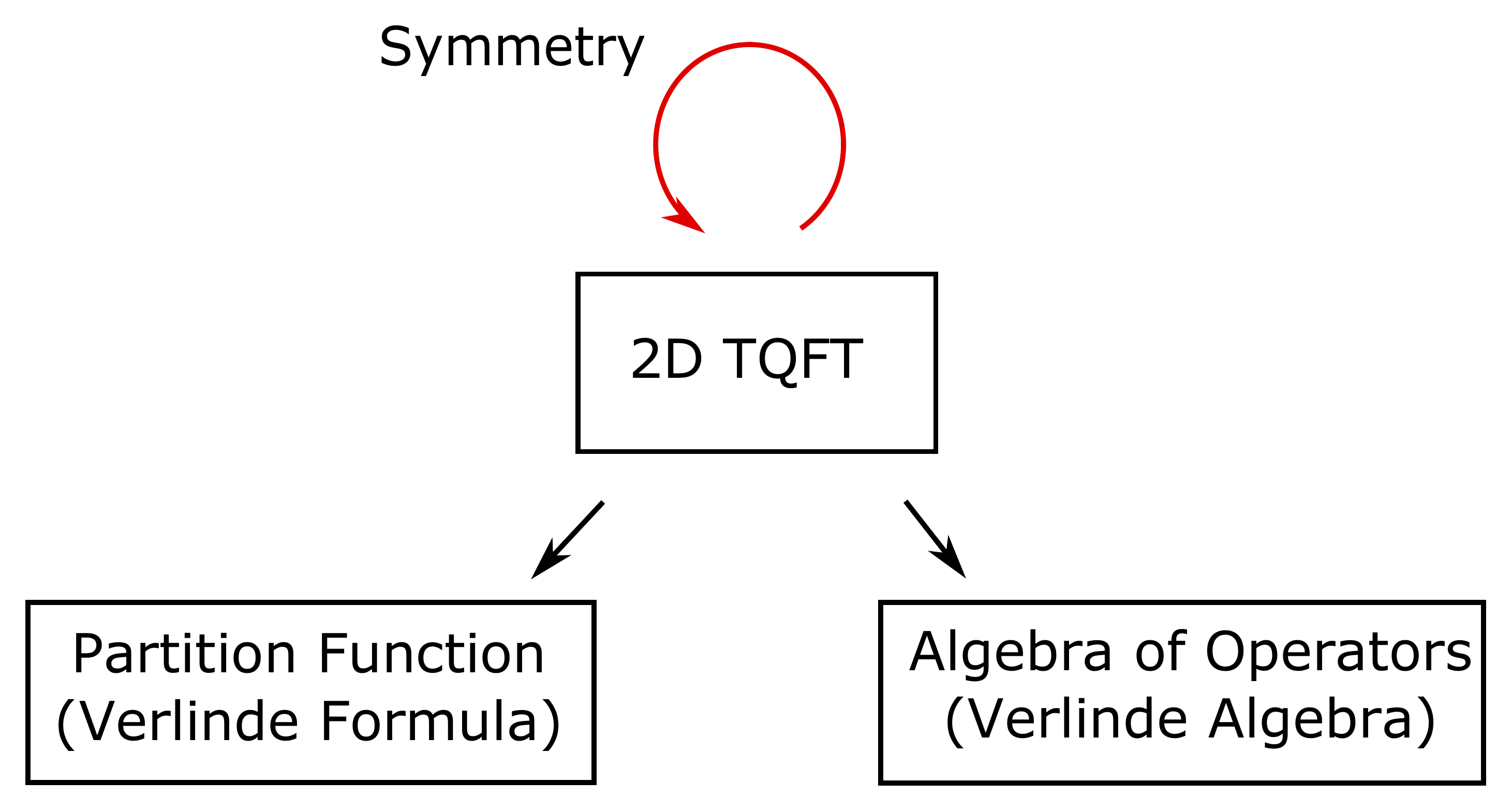}
    \caption{Symmetry of a 2D TQFT is reflected in its partition function as well as its algebra of local operators.}
    \label{fig:2DTQFT}
\end{figure}

2d TQFTs are ubiquitous in mathematics and physics, and they often serve as bridges between different subjects and fields. One classical example is the Verlinde formula \cite{VERLINDE1988360}, which not only counts conformal blocks of certain rational conformal field theories known as Wess--Zumino--Witten models, gives the dimension
of the space of generalized theta functions on a Riemann surface, but also enjoys a hidden connection with the quantum cohomology of Grassmannians \cite{Witten:1993xi} (see also \cite{cmp/1104248305}). To see this hidden connection, one will need to realize that the Verlinde formula is in fact a partition function of a two-dimensional topoogical quantum field theory (2d TQFT), which is also equipped with an algebra of local operators, also known as the Verlinde algebra. This is illustrated in Figure \ref{fig:2DTQFT}.

Another benefit of working with the fully-fledged 2d TQFT, instead of just its partition function, is that 2d TQFTs can have symmetries, which not only impose strong constraints on the algebraic structures that the TQFT encodes, but also enable one to produce families of distinct TQFTs by gauging these symmetries.

As we will see in this paper, such perspective naturally unifies different versions of the Verlinde formula associated with the same Lie algebra, and explain several observations made in the mathematical study of the generalized theta functions, including an observation by Beauville that there are variants of the Verlinde formula which are not directly associated with conformal field theories. To achieve such unification at the maximal extent, one will have to incorporate generalized symmetries such as 1-form and $(-1)$-form symmetries.

In this paper, we will systematically study generalized symmetries and their gauging in a class of 2d TQFTs that are ``semisimple,'' which obeys the original Atiyah--Segal axioms for TQFTs.\footnote{This is to ensure that we won't run into potential pathologies that sometime are present in theory that are topological in the boarder sense.  One example is the 2d Yang--Mills theory in the zero coupling limit (see e.g.~\cite{cmp/1104248198,Witten:1992xu} for a detailed study of this theory), which has an infinite-dimensional Hilbert spaces. However many results in the present paper still apply to these cases.} Such approach via generalized symmetries is the most useful when the 2d TQFTs come in family which all enjoy the same symmetry. This is indeed the case for the Verlinde formula, for which there exists a 1-parameter deformation known as the equivariant Verlinde formula \cite{Gukov:2015sna, Andersen:2016hoj, halpernleistner2016equivariant} that comes from a 1-parameter family of 2d TQFTs. They encode very interesting information about the geometry of the moduli space of \textit{Higgs} bundles on a Riemann surface,  just like the original formula can be used to probe the geometry of the moduli space of bundles. And one of the main goal of this paper is to apply the general results about symmetries of 2d TQFTs to investigate the problem of quantization of the moduli space of $G$-Higgs bundles when $G$ is not necessarily simply connected.

\subsection{Generalized symmetries of 2D semisimple TQFTs} \label{sec:symmetries_intro}

Global symmetry is one of the most important tools in the study of quantum field theories. In the past decade, the classical notion of symmetries has been generalized to the case of ``p-form" symmetries \cite{Gaiotto:2014kfa}. These symmetries share many of their properties with ordinary symmetries known to the practitioners of QFT since the twentieth century. For example, they can have anomalies, and when non-anomalous they can be gauged. Subsequent years have seen a vast amount of applications of these new symmetries in improving our understanding of strongly coupled systems.

Our focus in this paper will be on a class of QFTs in two dimensions known as ``semisimple topological QFTs". These QFTs are topological in the sense that their observables depend only on the topology of the underlying spacetime manifold. The algebra of the local operators in such a topological theory will turn out to be a commutative Frobenius algebra, and a well-known result regarding 2d TQFTs is that the information they contain is fully captured by the commutative Frobenius algebra,
\begin{equation}
    \text{2d TQFTs}\quad \Longleftrightarrow\quad \text{commutative Frobenius algebras}.
\end{equation}
Taking the product of two operators is sometimes referred to as ``fusion,'' while the bilinear form (also known as the Frobenius form, or, sometimes, the ``metric'') is defined by the $S^2$-partition function with insertion of two local operators.

In this paper we will focus on 2d TQFTs that are semisimple, in the sense that the corresponding commutative Frobenius algebra is semisimple. 

This is equivalent to the condition that this algebra has a basis $\CB =\{e_i\}$, sometime referred to as the ``idempotent basis,'' such that the fusion is diagonalized, $e_i e_j = \delta_{ij} N_j e_j$; and the pairing is orthonormal, $(e_i, e_j) = \delta_{ij}$. Such a basis is subject to the ambiguity of $e_j\mapsto -e_j$ for each $j$, which can be fixed by assuming that the  $N_j$'s all have positive real part.\footnote{$N_j$'s are often referred to as ``fusion eigenvalues'' or ``quantum dimensions.'' When the TQFT is unitary, $N_j$ take real values and we can further assume that they are positive\cite{Durhuus:1993cq}.} 

A global symmetry of a two-dimensional QFT is generated by a topological defect of dimension 0, 1 or 2, which corresponds respectively to a $1$-form, a $0$-form or a $(-1)$-form symmetry. While $(-1)$-form symmetries are not symmetries in the traditional sense as they alter the theory, 0- and 1-form symmetries must act in a very restricted way. We will show in this paper that for a semisimple TQFT,
\begin{itemize}
\item a 0-form symmetry necessarily acts as a permutation on $\CB$, permuting the basis elements with the same fusion eigenvalue $N_i$, and
\item a 1-form symmetry acts on the elements of $\CB$ by multiplication with a phase. 
\end{itemize}
These statements have interesting consequences. For example, it follows from the first statement that the action of any connected Lie group on a 2d semisimple TQFT must be trivial. Furthermore, if a 2d TQFT has $m$ identical eigenvalues, it will have an $S_m$ symmetry, and any symmetry of a 2d TQFT must be enhanced to a product of permutation groups.

Gauging generalized symmetries is often a useful way of constructing new QFTs. In particular, gauging a $p$-form symmetry of a semisimple TQFT produces another semisimple TQFT. So we can ask how the data of the new TQFT is determined in terms of the data of the original TQFT. In this paper, we give an algorithm to fully determined the commutative Frobenius algebra of the resulting theory from the data of the original theory. 

For example, consider the case of gauging a $\Z_n^\zero$ symmetry of a TQFT $\CT$ with diagonal basis $\CB$. We find that the gauged TQFT has a diagonal basis $\{e_i^{(a)}\}$ with $i \in\CB/\Z_n$ --- the set of all orbits of $\CB$ under the action of $\Z_n$, and $a \in \{1,2,\cdots, |\Stab(i)|\}$ --- the stabilizer of a representative of an orbit $i$. The fusion of $\{e_i^{(a)}\}$ is
\bea 
e_i^{(a)} e_j^{(b)} = \delta_{ij} \delta^{(a)(b)} \left(\frac{|\Stab(i)|}{n^\half}  N_i\right) e_i^{(a)}.
\eea
Note that the fusion eigenvalues are identical for each fixed $i$. Therefore, the gauged TQFT has a product 0-form symmetry, $\prod\limits_{i}S_{m_i}$ for $m_i = |\Stab(i)|$. In particular, the familiar ``dual" $\Z_n^\zero$ symmetry of the gauged TQFT is necessarily enhanced to this much larger group, and this fact is true about any semisimple TQFT.

\subsection{Quantization of moduli space of Higgs Bundles}
As an application of the general discussion of finite symmetries of two-dimensional TQFTs and their gaugings, we study the problem of quantization of the moduli space of Higgs bundles in this paper. 
This problem has already been studied in the recent years in the case of moduli space $\CM:= \CM(\widetilde{G},\Sigma)$ of $\widetilde{G}$-Higgs bundles on a genus-$g$ Riemann surface $\Sigma$ when $\widetilde{G}$ is either simply connected, or with a free $\pi_1$, such as $SU(N)$ or $U(N)$ \cite{Gukov:2015sna, Gukov:2016lki, Andersen:2016hoj, halpernleistner2016equivariant}. The moduli space is non-compact, and the procedure of geometric quantization leads to an infinite-dimensional Hilbert space identified with $H^0(\CM, \CL^k)$, the space of holomorphic sections of the  ``pre-quantum line bundle'' $\CL^k$. However, there exists a $\C^\star$-action on $\CM$ that lifts to an action on this infinite-dimensional space,
giving a decomposition
\bea \label{eq:decomposition_intro}
H^0 (\cM, \cL^k) = \oplus_{n=0}^{\infty} H^0_n (\cM , \cL^k).
\eea
The $\C^\star$-character, 
\bea 
\text{dim}_t \ H^0 (\cM, \cL^k) = \sum\limits_{n=0}^{\infty} \text{dim} \ H^0_n (\cM, \cL^k) t^n,
\eea
also known as the \textit{Hitchin character}, was computed in \cite{Gukov:2015sna, Andersen:2016hoj, halpernleistner2016equivariant}. When $\Sigma$ is a closed surface of genus $g>1$, it takes the value,
\bea 
\text{dim}_t H^0 (\cM, \cL^k) = \sum\limits_{\lambda \in P_k} \theta_t (f_{\lambda,t})^{1-g}.
\eea
The quantities appearing on the right hand side were defined in \cite{Andersen:2016hoj, halpernleistner2016equivariant}, and are reviewed in Section \ref{sec:review_simply_conn}. It was proved in \cite{Andersen:2016hoj} that the Hitchin character is the partition function of a two dimensional semisimple TQFT, which we denote by $\CT(\tG_k)$.

In this paper, we will study the quantization of the moduli space $\cM(G,\Sigma)$ of $G$-Higgs bundles when $G$ is not simply-connected. Let $\pi_1(G)=Z$, then these moduli spaces are identified with the quotient of $\CM$ by $H_1 (\Sigma, Z)$. This action and the $\C^\star$-action on $\CM$ commute, thus it leads to a natural question: what are the Hitchin characters $\text{dim}_t \ H^0 (\cM(G), \cL^k)$? The approach used in \cite{Andersen:2016hoj, halpernleistner2016equivariant} cannot be directly applied to this case (at least not without any modifications). On the other hand, $H^0(\CM(G),\CL^k)$ is identified with the subspace of sections in $H^0 (\cM, \cL^k)$ that are invariant under the action of $\CG:=H_1 (\Sigma, Z)$, giving
\bea \label{eq:quotient}
\text{dim}_t \ H^0 (\cM (G) , \cL^k) = \frac{1}{|\CG|} \sum\limits_{h \in \CG} \Trt (h| H^0 (\cM  , \cL^{k})).
\eea
Thus the problem of quantizing $\cM(G)$ is exchanged with the problem of finding the \textit{equivariant trace}, $\Trt (h| H^0 (\cM  , \cL^{k})) = \sum\limits_{n=0}^{\infty} \Tr (h| H^0_n (\cM  , \cL^{k})) t^n$ for each $h \in \CG$. 

Our understanding of finite symmetries and their gaugings presented in Section \ref{sec:symmetries_intro} helps solve this problem. In particular, the TQFT $\CT(\widetilde{G}_k)$ has the center symmetry as (a subgroup of its) 0-form and 1-form symmetry groups. We argue in Section \ref{sec:quantization} that the equivariant trace $\Trt (h| H^0 (\cM  , \cL^{k}))$ is equal to the partition function of $\CT(G_k)$ in the presence of certain 0-form and 1-form symmetry backgrounds. Here we will present some examples. 
 
Let us consider the simplest case $\tG = SL(2,\C)$, $k\equiv 0 \pmod 4$ and a non-zero $h\in H_1(\Sigma, \Z_2)$.\footnote{The reason we restrict to $k \equiv 0 \pmod 4$ is explained in Section \ref{sec:review_simply_conn}. See the discussion around equation \eqref{eq:transgressive}.} We find
 \bea 
\Trt (h| H^0 (\cM  , \cL^{k})) = \left( \frac{\frac{k}{2}+1 +\left( \frac{k}{2}-1 \right) t}{ (1-t)(1+t)^3} \right)^{g-1}.
\eea
This readily gives the Hitchin characters for $\cM(PSL(2,\C))$ using \eqref{eq:quotient}. For another example, consider the case $\widetilde{G} = SL(4,\C)$ and $k \equiv 0 \pmod 8$. If $\cM_d \equiv \cM_d(SL(4,\C))$ denotes the moduli space of semi-stable $SL(4,\C)$-Higgs bundles of degree $d$ (where $d \in \{0,1,2,3\}$), we find that 
for any non-zero $h \in H_1(\Sigma, \Z_2) \subset H_1(\Sigma, \Z_4)$,
\bea \label{eq:tr_sl4_intro}
\Trt  (h|H^0 (\cM_d, \cL^k)) =
 \sum\limits_{\substack{\lambda = (a, \frac{k}{2}-a, a)  \\ a\leq \frac{k}{2}} } 
 (-1)^{ad} \theta_t (f_{\lambda, t}).
\eea 
The quantities appearing on the right-hand side are defined in Section \ref{sec:verlinde_simple}. Here we remark that it depends only on the parity of  $d$, and not on its actual value. This is a surprising result as the geometry of the moduli space $\CM_d$ for different values of degree $d$ are vastly different, but our results show that the equivariant traces some times turn out to be the same.

Using the formulae for equivariant trace, we will find the Hitchin characters $\text{dim}_t \ H^0 (\cM (G) , \cL^k)$ for an arbitrary complex simple  Lie group $G$. 
For example, when $G=PSL(2,\C)$, and the underlying surface $\Sigma$ is a genus-$2$ curve, we find 
\bea 
\text{dim}_t \ H^0 (\cM_{w_2=0}(PSL(2,\C)) , \cL^k) = \frac{1}{(1-t)^3} \left(c_3 k^3 + c_2 k^2 + c_1 k + c_0\right)
\eea
where 
\bea \nonumber
c_3 &=& \frac{1}{96}, \\ \nonumber
c_2 &=& \frac{1}{16}\frac{1+t^2}{1-t^2}, \\ \nonumber
c_1 &=& \frac{7-27t + 34 t^2 - 27 t^3 + 7t^4}{12(1-t^2)^2}, \\ \nonumber
c_0 &=& \frac{1-16t + 15 t^2 - 19 t^3 + 15 t^4 - 6t^5 + t^6 - t^{k+3}}{(1-t^2)^3}.
\eea

Finally, we remark that as in \cite{Gukov:2015sna}, the zeroth piece in the decomposition \eqref{eq:decomposition_intro} (and its analogue in the non-simply-connected case) is identified with the Hilbert space of compact group Chern--Simons theory on $\Sigma$. The $t\to 0$ limit of our formulae give quantization of the moduli space of semi-stable holomorphic $G$-bundles. In special cases, these formulas have been obtained in the literature using more geometric methods \cite{10.1215/S0012-7094-94-07618-7, beauville1996verlinde, meinrenken2018verlinde}. In each case, the $t\to 0$ limit of our results matches with the formulae found there.

\subsection{Quantization of moduli space of parabolic Higgs bundles}

We also discuss the quantization of the moduli space of parabolic semi-stable $G$-Higgs bundles in this paper.\footnote{When the genus is small, the right geometric interpretation of the equivariant Verlinde formula is often in terms of the moduli \emph{stack} of (parabolic) Higgs bundles. However, we will not be concerned with this distinction in the present paper. For more details, see e.g.~Section~5 in \cite{Andersen:2016hoj}.} When $G= \tG$, i.e.~the group is simply-connected, the quantization of these spaces was discussed in \cite{Gukov:2015sna, Gukov:2016lki, Andersen:2016hoj}. The Hitchin character is captured by the partition functions of $\CT(\tG_k)$ on punctured surfaces. In particular, $\CT(\tG_k)$ has a distinguished basis called the ``parabolic basis" in which the fusion rules and the metric compute the Hitchin characters for the moduli spaces of parabolic Higgs bundles on three-punctured and two-punctured spheres respectively. The algebra given by the fusion rules of $\CT(\tG_k)$ in this basis is sometimes referred to as the ``equivariant Verlinde algebra"  \cite{Gukov:2015sna}.

As an example, consider the case of $\tG=SL(2,\C)$ where a choice of parabolic structure is determined by an integrable weight $j\in\{0,1,\cdots, k\}$. As we will review in greater detail in Section \ref{sec:PSL2Calgebra}, in the corresponding parabolic basis, $\CT(SL(2,\C)_k)$ assigns
\bea 
f^{a_1 a_2 a_3} = 
\begin{cases} 
1 &\mbox{if }a_1 + a_2 + a_3 \in 2\mathbb{Z} \ \text{and} \Delta \leq 0 \\
t^{\Delta /2} &\mbox{if } a_1 + a_2 + a_3 \in 2\mathbb{Z} \ \text{and} \Delta  > 0 \\
0 & \mbox{otherwise} 
\end{cases}
\eea 
to the three-punctured sphere labelled by $a_i$. This is also the Hitchin character of the moduli space of semi-stable parabolic $SL(2,\C)$-Higgs bundles on the surface \cite{Gukov:2015sna, Andersen:2016hoj}. 

In this paper, we extend the results of \cite{Gukov:2015sna, Gukov:2016lki, Andersen:2016hoj} to the case when $G$ is not simply-connected. Our main example in Section \ref{sec:PSL2Calgebra} is that of $G=PSL(2,\C)$. 
The corresponding TQFT $\CT(PSL(2,\C)_k)$ is obtained by gauging a $\Z_2^\zero \times \Z_2^\one$ symmetry of $\CT(SL(2,\C)_k)$. We propose that there exists a distinguished basis of $\CT(PSL(2,\C)_k)$ in which the fusion rules and the Frobenius bilinear form compute the Hitchin characters for the moduli space of parabolic $PSL(2,\C)$-Higgs bundles on three-punctured and two-punctured spheres respectively. 

An interesting fact that might be surprising from the purely geometric point of view is that  $\CT(PSL(2,\C)_k)$ now has an extra ``twisted state,'' which has to be included to fully characterize the 2d TQFT. In other words, the corresponding Frobenius algebra now has basis elements,
\bea \label{eq:spectrum_intro}
x^0 , x^2, \cdots, x^{\frac{k}{2}-2}, x^{{\frac{k}{2}}^\one}, x^{{\frac{k}{2}}^{(2)}},
\eea
where the superscript denotes an integrable highest weight of $SL(2,\C)$. The appearance of an additional operators labeled by the highest weight $\frac{k}{2}$ is essential to ensure that the Hitchin characters of $\cM(PSL(2,\C))$ is well behaved under cutting and gluing.  It will be interesting to better understand its geometric interpretation in terms of the moduli space of Higgs bundles, which we hope to further investigate in future work. 

We check the physics prediction against geometry in Section \ref{sec:parabolic_torus}, where we compute the Hitchin character for $\cM(PSL(2,\C))$ in the case of once-punctured torus in two ways. We carry out this computation first through TQFT methods, and then through a Lefschetz trace formula due to Atiyah and Singer \cite{10.2307/1970717}, and show that the two match. In either case, we find
\bea 
\Trt (h|H^0 (\cM_{d}, \cL^k)) = \frac{(-1)^{a/2}}{1+t}
\eea
which gives the following expressions for the Hitchin character,
\bea \nonumber 
\text{dim}_t H^0 (\cM_{w_2=0}(PSL(2,\C), \cL^k) &=& \frac{1}{4}\left( \frac{k-a+1}{1-t} + \frac{2t}{(1-t)^2} + \frac{4t^{a/2}}{(1-t^{-1})(1-t^2)} +\frac{3 (-1)^{a/2}}{1+t}\right)
\eea
and 
\bea  \nonumber
\text{dim}_t H^0 (\cM_{w_2=1}(PSL(2,\C), \cL^k) &=& \frac{1}{4}\left( \frac{a+1}{1-t} + \frac{2t}{(1-t)^2} + \frac{4t^{(k-a)/2}}{(1-t^{-1})(1-t^2)} +\frac{3 (-1)^{a/2}}{1+t}\right).
\eea

\section{Finite gauging of 2d TQFTs}\label{sec:FiniteGauging}
In this section, we will discuss finite symmetries of two-dimensional semisimple TQFTs and study their gauging. In Section \ref{sec:SymmetriesFrobenius}, we will discuss the action of 0-form and 1-form symmetries on the TQFT in the  basis that diagonalizes the fusion rule. We will show that a 0-form symmetry necessarily acts as a permutation on the basis elements, and that a 1-form symmetry acts by multiplication by a root of unity. In Section \ref{sec:gauging}, we will study the gauging of abelian 0-form and 1-form symmetries. As a result, we will find the partition functions and the Frobenius algebras of the new theories obtained after gauging.

\subsection{Symmetries of two-dimensional TQFTs} \label{sec:SymmetriesFrobenius}
\begin{figure}[!ht]
    \subfloat[The action of 0-form symmetry\label{subfig-1:0formonH}]{%
      \includegraphics[width=0.45\textwidth]{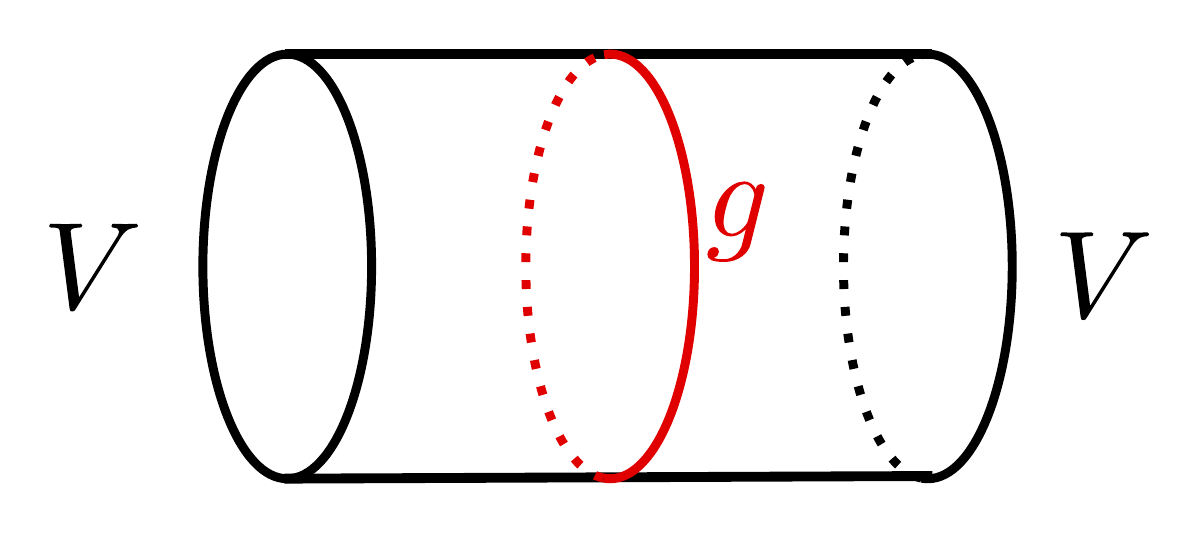}
    }
    \hfill
    \subfloat[The action of 1-form symmetry\label{subfig-2:1formonH}]{%
      \includegraphics[width=0.45\textwidth]{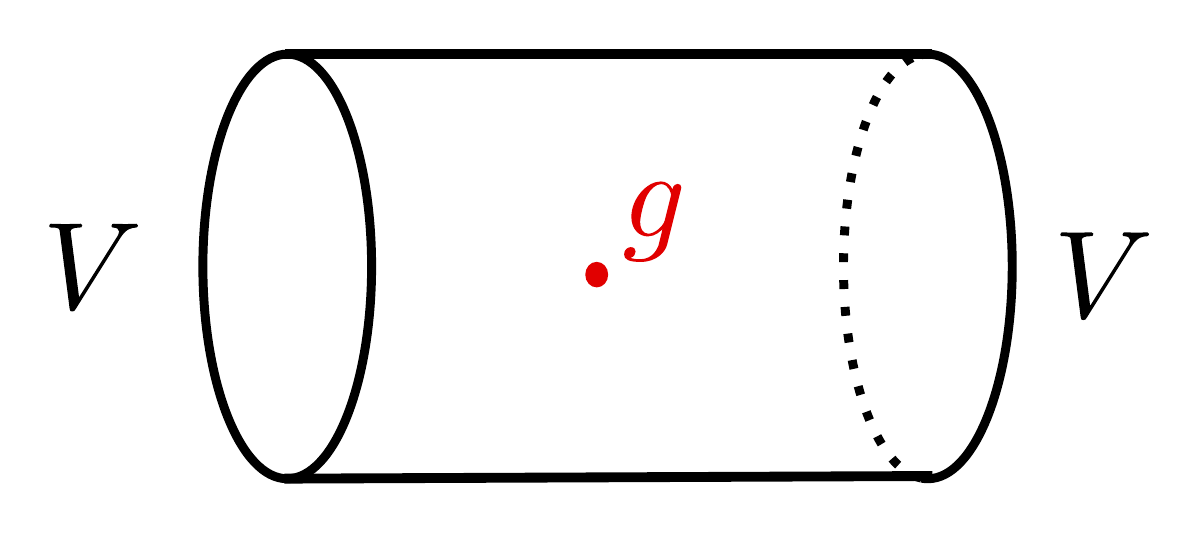}
    }
    \caption{The action of 0-form and 1-form symmetry transformations on a state in Hilbert space.}
    \label{fig:actiononH}
  \end{figure}

\begin{figure}[!ht]
    \subfloat[The action of 0-form symmetry\label{subfig-1:0form}]{%
      \includegraphics[width=0.45\textwidth]{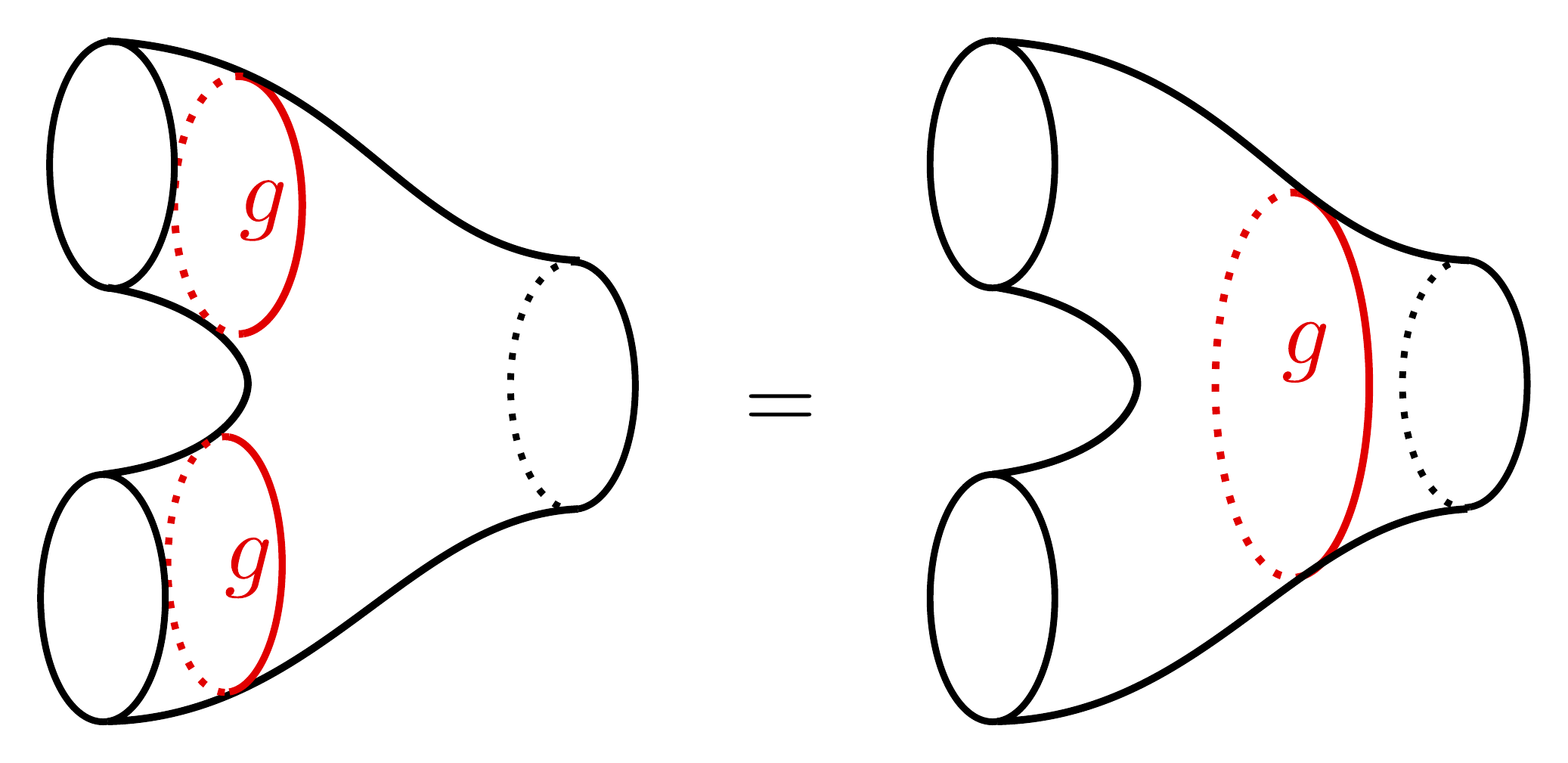}
    }
    \hfill
    \subfloat[The action of 1-form symmetry\label{subfig-2:1form}]{%
      \includegraphics[width=0.45\textwidth]{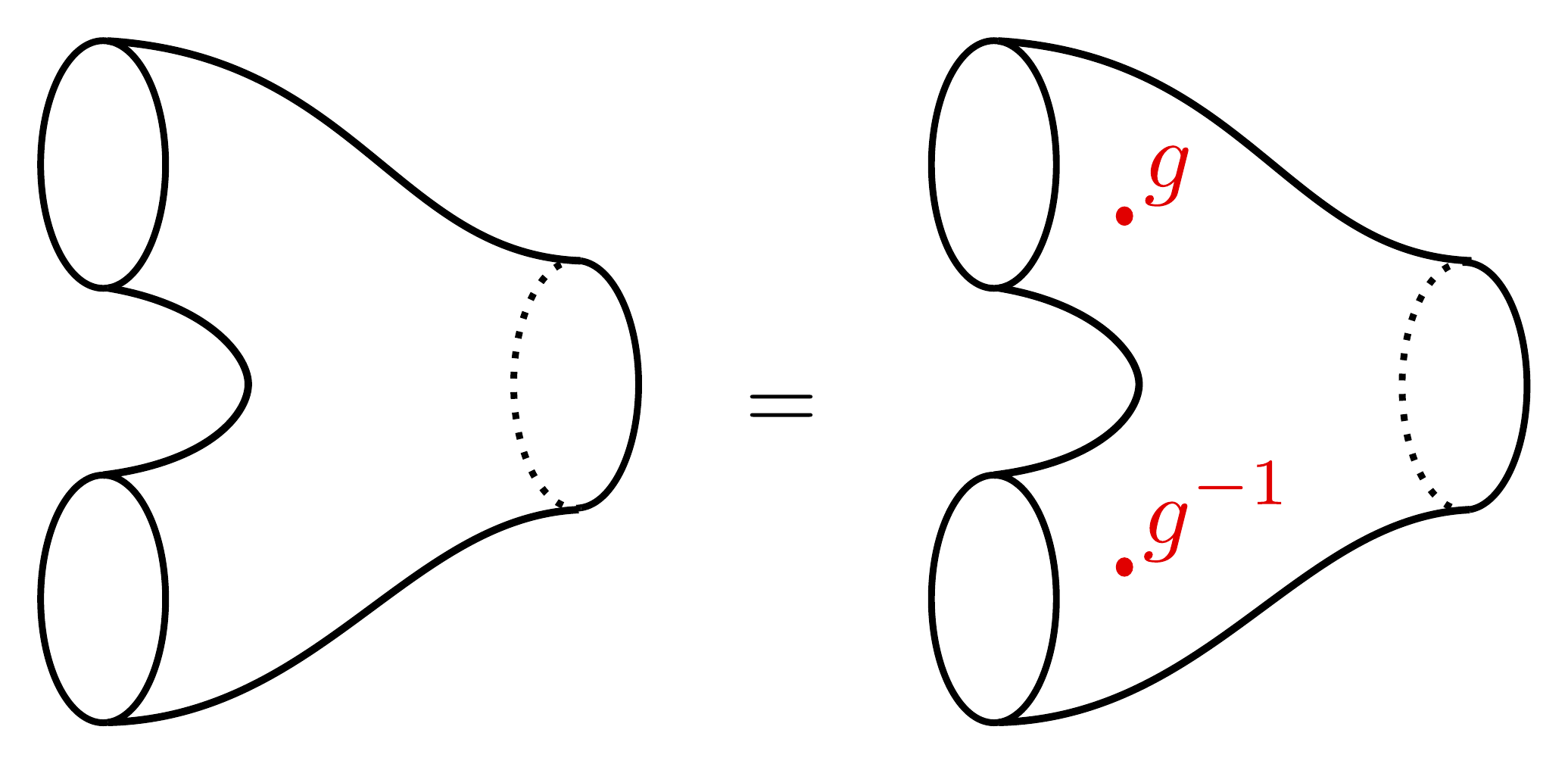}
    }
    \caption{The action of 0-form and 1-form symmetry transformations on surfaces in a two-dimensional TQFT.}
    \label{fig:action}
  \end{figure}

A symmetry of a two-dimensional QFT is generated by a topological defect of dimension 0, 1, or 2 which respectively corresponds to 1-form, 0-form and $(-1)$-form symmetries. Among these, both 1-form and 0-form symmetries can act on the Hilbert space on $S^1$. This is depicted in Figure \ref{fig:actiononH}. In the language of commutative Frobenius algebras, having a 0-form symmetry means that the bilinear form is invariant 
\begin{equation}
    (v_1,v_2)=(g\cdot v_1,g\cdot v_2),
\end{equation}
where $g$ is a 0-form symmetry transformation, and that the fusion of two operators satisfies
\begin{equation}
    (g\cdot v_1) \times (g\cdot v_2)=g\cdot(v_1\times v_2).
\end{equation}
Similarly, when $g$ is a 1-form symmetry transformation, the bilinear form is invariant in the sense that
\begin{equation}
    (v_1,v_2)=(g\cdot v_1,g^{-1}\cdot v_2),
\end{equation}
and the fusion product satisfies
\begin{equation}
    (g\cdot v_1) \times (g^{-1}\cdot v_2)=v_1\times v_2.
\end{equation} 
The geometric meanings of these equalities are illustrated in Figure \ref{fig:action}.
  
A ``semisimple'' 2d TQFT---meaning the commutative Frobenius algebra is semisimple---can be decomposed into ones with one-dimensional Hilbert space. This amounts to diagonalizing the Frobenius algebra into a direct sum of one-dimensional subalgebras. In other words, there exists a basis $\CB = \{e_i\}$ that diagonalizes both the pairing $(e_i,e_j)=\delta_{ij}$ and the fusion $e_i e_j = \delta_{ij} N_i e_i$. The trace map sends $e_i$ to $\frac{1}{N_i}$. We will refer to $N_i$'s as the``fusion eigenvalues,'' and write the decomposition as $\CA=\bigoplus_{i}\C_{N_i}$.\footnote{\label{footnote_choice}There is an ambiguity in the choice of a diagonal basis that satisfies the stated conditions: $\{-e_i\}$ gives another basis but with eigenvalues $-N_i$. We will fix this ambiguity by imposing $\text{arg}(N_i)\in [0,\pi)$.}

We will now proceed to show that 
\begin{itemize}
    \item a 0-form symmetry transformation acts on the set of $e_i$'s by permutation where it can only permute the elements with the same fusion eigenvalue, and
    \item a 1-form symmetry acts on $e_i$ by a phase.
\end{itemize}

To see the first, assume that a 0-form symmetry transformation labeled by $g$ acts on $e_i$ as
\begin{equation}
    g\cdot e_i=\sum_j c_{ij}e_j.
\end{equation}
As symmetry preserves the inner product,
\begin{equation}
    (g\cdot e_i,g\cdot e_\ell)=\delta_{i\ell},
\end{equation}
so we have
\begin{equation}\label{0FormConstraint1}
    \sum\limits_j c_{ij}c_{\ell j}=\delta_{i\ell}.
\end{equation} 
In other words, $c_{ij}$ can be assembled into an orthogonal matrix. 
On the other hand, we also have the equality
\begin{equation}
    (g\cdot e_i) \times (g\cdot e_\ell)=g\cdot(e_i \times e_\ell),
\end{equation}
which leads to
\begin{equation}\label{0FormConstraint}
c_{ij}c_{\ell j}  N_j=\delta_{i\ell}c_{ij} N_i.
\end{equation}
To see what this equation implies, first choose $\ell=i$. Then we have 
$    N_jc_{ij}^2=N_ic_{ij}$,
which tells us that
\begin{equation}
    c_{ij}=0 \text{ or } N_i/N_j.
\end{equation}
But for $\ell\neq i$, Equation~\eqref{0FormConstraint} becomes
\begin{equation}
    c_{ij}c_{lj}=0
\end{equation}
for all $j$. Therefore, if a $c_{ij}$ is non-zero, it will be the only non-zero entry in the $j$-th column. Furthermore, it also has to be the only non-zero entry in the $i$-th row, as $c^{T}$ represents the action of $g^{-1}$. Since the choice of $j$ is unique given $i$, we can define a function $\sigma$ such that $j=\sigma(i)$. It will be single-valued and hence is a permutation. The constraint \eqref{0FormConstraint1} now implies that $N_i=N_j$ if $j=\sigma(i)$.\footnote{
Note that \eqref{0FormConstraint1} really says $N_i^2 = N_j^2$ if $j=\sigma(i)$. However, the other possibility $N_i=-N_j$ is not consistent with footnote \ref{footnote_choice}, i.e. $\text{arg}(N_i) \in [0,\pi)$.} 
Therefore, the action of $g$ is by permutation $g\cdot e_{i}=e_{\sigma(i)}$ and it can only permute $e_i$'s with the same fusion eigenvalues. 

This simple fact has many interesting consequences. For example, the action of any connected Lie group on a 2d TQFT has to be trivial. On the other hand, if a 2d TQFT has $m$ identical eigenvalues, it will have an $S_m$ symmetry, and any symmetry group of a 2d TQFT will be enhanced to a product of permutation groups.

\smallskip

Now we show the second point mentioned above, i.e. the action of 1-form symmetry on the elements in the diagonal basis must be as multiplication by a phase. Let $g$ be a 1-form symmetry transformation, then 
\begin{equation}
    (g\cdot e_i) \times e_j=e_i \times (g\cdot e_j)
\end{equation}
tells us that $g \cdot e_i = \chi_i (g) e_i$ for some $\chi_i(g)  \in \C^\times$. When $g$ is an element of finite order, $\chi_i (g)$ is in fact a root of unity.

\smallskip

\subsection{Gauging of 0-form and 1-form symmetries} \label{sec:gauging}
In this subsection we will consider gauging abelian 0-form and 1-form symmetries. A finite abelian group is a product of cyclic groups so we may gauge one factor at a time. Therefore, we will focus on gauging a $\Z_n^\zero$ and a $\Z_n^\one$ symmetry of a TQFT $\CT$. As a result we will find the partition functions and the commutative Frobenius algebras associated to the TQFTs $\CT/\Z_n^\zero$ and $\CT/\Z_n^\one$.\footnote{We will assume throughout this section that these symmetries are non-anomalous. Of course, if these symmetries were anomalous, they could not be gauged.}

Let us first start with the case of $\CT/\Z_n^\zero$. The partition function $\CZ_{\CT/\Z_n^\zero}$ on a closed surface $\Sigma$ of genus $g$ is obtained by summing over principal $\Z_n$ bundles. These are parameterized by elements of $H_1(\Sigma, \Z_n)$, and we write\footnote{Our choice of normalization factor is such that gauging the $\Z_n^\zero$ symmetry twice gives back the partition function of $\CT$ on the nose, and not up to the partition function of an Euler TQFT.}
\bea 
\CZ_{\CT/\Z_n^\zero} = \frac{1}{|H^1(\Sigma, \Z_n)|^{\frac{1}{2}}} \sum\limits_{h \in H_1 (\Sigma, \Z_n)} \CZ_{h}.
\eea
For $h = 0$, $\CZ_h$ is the ordinary partition function of $\CT$. It is well known that in terms of the fusion eigenvalues $N_i$ introduced in the previous subsection,
\bea \label{eq:Z_orig}
\CZ_{\CT} = \sum\limits_{i \in \CB} N_i^{2g-2}.
\eea 
We will show in Appendix \ref{sec:part_func} that
\bea \label{eq:Z_main}
\CZ_{\CT/\Z_n^\zero} &=& \sum\limits_{i \in \CB/\Z_n}|\Stab(i)| \left( \frac{|\Stab(i)|}{|\Z_n|^\half}  N_i \right)^{2g-2}
\eea
where $\CB/\Z_n$ is the set of all orbits of $\CB$ under the $\Z_n$ action. For an orbit $i$, we denote the cardinality of the stabilizer of any of its representatives in $\CB$ as $|\Stab (i)|$. A crucial ingredient in the derivation of this result will be the partition function $\CZ_h$ which we also compute in Appendix \ref{sec:part_func},
\bea \label{eq:Z_h}
\CZ_{h} = \sum\limits_{i \in \CB^{\Z_m}}  N_i^{2g-2}.
\eea
Here $\Z_m$ is the smallest subgroup of $\Z_n$ such that $H_1(\Sigma, \Z_m)$ contains $h$, and $\CB^{\Z_m}
\subset \CB$ is the subset of fixed points of $\Z_m$. 

The commutative Frobenius algebra associated to the gauged TQFT $\CT/\Z_n^\zero$ can be read off from equation \eqref{eq:Z_main}. This algebra has a diagonal basis $\{ e_i^{(a)}\}$ for $i \in \CB/\Z_n$ and $a\in |\Stab(i)|$. The fusion of these operators is 
\bea 
e_i^{(a)} e_j^{(b)} = \delta_{ij} \delta^{(a)(b)} \left(\frac{|\Stab(i)|}{|\Z_n|^\half}  N_i\right) e_i^{(a)}.
\eea
We note that the fusion eigenvalues are independent of the value of $a$, i.e. the gauged TQFT has a $\prod\limits_{i}S_{m_i}$ symmetry for $m_i = |\Stab(i)|$. This is an enhancement of the familiar ``dual" $\Z_n^\zero$ symmetry obtained after gauging a $\Z_n^\zero$ symmetry. We emphasize that this enhancement occurs for \textit{all} two-dimensional semisimple TQFTs. 

We turn now to the case of $\CT/\Z_n^\one$. As discussed in Section \ref{sec:SymmetriesFrobenius}, a 1-form symmetry transformation acts on $\CB$ by multiplication with roots of unity, $z. e_i = \chi_i (z) e_i$. Here, $z \in \Z_n$ denotes a local operator in $\CT$. As $\Z_n$ is gauged, an operator $e_i$ with $\chi_i (z) \neq 1$ for any $z \in \Z_n$ is projected out. 
The diagonal basis of $\CT/\Z_n^\one$ is the subset of operators $\{e_i | \chi_i (z) = 1 \ \forall z \in \Z_n\}$ with fusion rules
\bea \label{eq:fusion_1_form}
e_i e_j = \delta_{ij} N_i n^{-\half} e_i.
\eea
The normalization factor $n^{-\half}$ has been fixed by comparison with the partition function that we now compute. 

The partition function of $\CT/\Z_n^\one$ is obtained by summing over various $\Z_n$ gerbes which are parameterized by $H^2(\Sigma, \Z_n) = \Z_n$. In particular,
\bea 
\CZ_{\CT/\Z_n^{(1)}} = \frac{1}{|H^1 (\Sigma, \Z_n)|^\half} \sum\limits_{z \in \Z_n} \CZ_z .
\eea
Here, $\CZ_z$ is the partition function of $\CT$ in the presence of the local operator $z$. The equation $z. e_i = \chi_i (z) e_i$ gives
\bea \label{eq:1form_w_background}
\CZ_z = \sum\limits_{i\in \CB} \chi_i (z) N_i^{2g-2}.
\eea
Summing over $\Z_n$ backgrounds, we obtain
\bea\label{eq:1formgauged}
\CZ_{\CT/\Z_n^{(1)}} =  \sum\limits_{\{e_i | \chi_i (z) = 1 \ \forall z \in \Z_n\}} \left(\frac{N_i}{|\Z_n|^\half} \right)^{2g-2}
\eea
which gives the result in \eqref{eq:fusion_1_form}. 

We will end this section with two remarks. The first concerns gauging $(-1)$-form symmetries. The presence of such a symmetry implies that there is a parameter $\theta$ and a family of theories $\CT_\theta$ parametrized by it \cite{Cordova:2019jnf,Cordova:2019uob}. Gauging a $(-1)$-form symmetry at the level of Frobenius algebra corresponds to taking direct products. One way that $(-1)$-form symmetry in 2d naturally arises is via gauging 1-form symmetries, where the $\theta$ parameter often undergoes by the name of ``discrete torsion" (see e.g.~\cite{Sharpe:2000ki} for a review, and also \cite{Hellerman:2006zs} and references therein for discussions about ``decomposition,'' which is a closely related phenomenon). 

The second comment is about gauging at the level of extended 2d TQFTs. These are theories in which one is allowed to have boundary conditions. Just as commutative Frobenius algebras are equivalent to 2d (non-extended) TQFTs, 2d extended TQFTs are fully described by Calabi--Yau categories (see \cite{Moore:2006dw} for a detailed review of this subject). As the simple objects of a Calabi--Yau category are in one-to-one correspondence with idempotent basis vectors that diagonalize the fusion rule, it is straightforward to translate the prescription of gauging at the level of Frobenius algebra to that at the level of Calabi--Yau categories.
 
\section{Equivariant Verlinde formula for any simple group} \label{sec:verlinde_simple}
In this section we will use the machinery developed in \autoref{sec:FiniteGauging} to compute the equivariant Verlinde formula for $\cM (G)$ --- the moduli space of semi-stable $G$-Higgs bundles for a connected simple complex Lie group $G$. We denote the universal cover of $G$ by $\tG$. As $\tG$ is simply connected, the answer for it is known \cite{Gukov:2015sna, Andersen:2016hoj, halpernleistner2016equivariant}, and is shown to be the partition function of a two-dimensional semisimple TQFT \cite{Gukov:2015sna, Andersen:2016hoj}. We will study symmetries of this TQFT and their gaugings. As a consequence, we will obtain the equivariant Verlinde formula for $\cM (G)$ for any simple Lie group $G$.

\subsection{Review: the simply connected case}  \label{sec:review_simply_conn}
Let $\tG$ be a simple and simply connected complex Lie group, and let $\cN$ be the moduli space of semi-stable $\tG$-vector bundles of degree zero on a genus-$g$ surface $\Sigma$. For $k \in \Z_{\geq 0}$, the Verlinde formula computes the dimension of $H^0 (\cN, \cL^k)$ where $\cL$ is the determinant line bundle over $\cN$. This formula is described as follows. 

We denote by $\tT$ a maximal torus of $\tG$ and by $W$ the Weyl group. Also, let $\mathfrak{t} \subset \mathfrak{g}$ be the Cartan subalgebra of $\mathfrak{g}$ associated to $\tT$. There exists a unique inner product $\langle - , - \rangle$ on $\mathfrak{g}$ such that the highest root has length squared equal to 2. If $h$ is the dual coxeter number of $\mathfrak{g}$ we define $i_k: \mathfrak{t} \to \mathfrak{t}^\star$ as
\bea 
i_k (a) = (k+h) \langle a, -  \rangle.
\eea
This map can be exponentiated to a map between tori,
\bea \label{eq:chik}
\chi_k: \tT \to \tT^\star.
\eea 
Now let $\rho = \frac{1}{2} \sum\limits_{\alpha \in \mathfrak{R}_+} \alpha $ be the Weyl vector defined with a choice of positive roots $\mathfrak{R}_+$, and consider the equation $\chi_k (f) = e^{2\pi i \rho}$. We denote by $\tT_{k}$ the set of all solutions to this equation, and by $\tT^{\text{reg}}_{k}$ the subset of regular solutions. We define
\bea 
\theta(f) = \frac{\Delta (f)^2}{|\tT_{k}|}
\eea
where
\bea \label{eq:theta}
\Delta = \prod\limits_{\alpha \in \mathfrak{R}_+} 2 \sin \left( \frac{i \alpha}{2} \right).
\eea 
The Verlinde formula computes the dimension of the space of sections of $\cL^k$ where $\cL$ is the determinant line bundle over $\cM$ \cite{teleman2009index},
\bea \label{eq:GVerlindeFormula}
\text{dim} \ H^0 (\cM_v, \cL^k) = \sum\limits_{f \in \tT^{\text{reg}}_{k}/W} \theta (f)^{1-g}.
\eea

There exists a bijection between the set of conjugacy classes $\tT^\reg_k/W$ and the set $P_k$ of highest integrable weights of $\tG$ at level $k$. The identification is $P_k \ni \lambda \to f_\lambda = e^{2\pi i (\lambda+ \rho) / (k+h)}$. The expression \eqref{eq:GVerlindeFormula} is equivalently written as\footnote{ See, for example, \cite{beauville1994conformal} for an argument.} 
\bea \label{eq:GVerlindeFormulainPk}
\text{dim} \ H^0 (\cM_v, \cL^k) = \sum\limits_{\lambda \in P_k} \theta (f_\lambda)^{1-g}.
\eea

\paragraph{} In \cite{Gukov:2015sna, Andersen:2016hoj, halpernleistner2016equivariant}, a similar formula was derived for the moduli space $\cM := \cM(\tG)$ of semi-stable $\tG$-Higgs bundles. It is well known that $\cM$ is non-compact, and hence its quantization $H^0 (\CM, L^k)$ must be infinite dimensional. However, $\cM$ admits a $\C^\star$-action which naturally lifts to a $\C^\star$-action on $\cL$, giving a decomposition
\bea 
H^0 (\cM, \cL^k) = \oplus_{n=0}^{\infty} H^0_n (\cM , \cL^k).
\eea
The character of $\C^\star$, also known as the \textit{Hitchin character}
\bea 
\text{dim}_t \ H^0 (\cM, \cL^k) = \sum\limits_{n=0}^{\infty} \text{dim} \ H^0_n (\cM, \cL^k) t^n
\eea
is given by a $t$-deformation of the Verlinde formula. In particular, we consider a deformation of the map \eqref{eq:chik},
\bea \label{eq:deformedBethe}
\chi_{k,t} = \chi_k \prod\limits_{\alpha \in \mathfrak{R}_+} \left( \frac{1-t e^\alpha}{1-t e^{-\alpha}}\right)^\alpha : \tT \to \tT^\star
\eea
and consider solutions to the equation $\chi_{k,t} (f) = e^{2\pi i \rho}$. Each solution can be expanded in powers of $t$ as $f_{0}+ t f_{1} + t^2 f_2 + \cdots$ where $f_0 \in \tT_{k}$. 
We denote the set of $t$-deformed solutions for which $f_0$ is regular as $\tT^{\text{reg}}_{k,t}$. Then the Hitchin character is given as
\bea \label{eq:GEqVerlindeFormula}
\text{dim}_t H^0 (\cM, \cL^k) = \sum\limits_{f \in \tT^{\text{reg}}_{k,t}/W} \theta_t (f)^{1-g}.
\eea
Here $\theta_t$ is a $t$-deformation of $\theta$ in \eqref{eq:theta},
\bea \label{eq:thetat}
\theta_t = \frac{(1-t)^{\text{rank}(G)}}{|\tT_{k}| \ \text{det} \ H_t^\dagger} \prod\limits_{\alpha} (1-e^{\alpha}) (1-t e^{\alpha})
\eea
where the product is taken over all roots, and $H_t^\dagger$ is the endomorphism of $\mathfrak{t}$ corresponding to the Hessian of the function $D_t : \mathfrak{t} \to \mathbb{C}$,
\bea 
D_t (a) = \frac{1}{2} (k+h) \langle a, a \rangle - \tr_{\mathfrak{g}} (\text{Li}_2 (te^{a})) + \langle \rho, a\rangle.
\eea
Equivalently we may use the bijection between $\tT^\reg_k/W$ and $P_k$ to write 
\bea \label{eq:GEqVerlindeFormulainPk}
\text{dim}_t H^0 (\cM, \cL^k) = \sum\limits_{\lambda \in P_k} \theta_t (f_{\lambda,t})^{1-g}
\eea
where $f_{\lambda,t} \in \tT^\reg_{k,t}/W$ satisfies $\lim_{t\to 0} f_{\lambda,t} = f_{\lambda}$.

Equation \eqref{eq:GEqVerlindeFormula} is what the authors of \cite{Gukov:2015sna} called the \textit{equivariant Verlinde formula}. It was argued in \cite{Gukov:2015sna,Andersen:2016hoj} that, for each level $k$, it is the partition function of a two-dimensional semisimple TQFT called the \textit{equivariant Verlinde TQFT}. As this TQFT depends on the choice of $\tG$ and $k$, we will denote it as $\CT (\tG_k)$. 
In the limit $t\to 0$, it reduces to the usual Verlinde TQFT which encodes the fusion algebra of Wilson lines of a compact group Chern--Simons theory. Its partition function \eqref{eq:GVerlindeFormula} computes the partition function of compact group Chern--Simons theory on $S^1 \times \Sigma$.

Similarly, $\CT(\tG_k)$ is also an $S^1$ reduction of a three-dimensional QFT --- ``the $\beta$-deformed complex Chern-Simons theory'' which is now only partially topological \cite{Gukov:2015sna} . This QFT can be obtained by having 3d $\cN=2$ Chern--Simons theory with an adjoint hypermultiplet on $S^1 \times \Sigma$, with a partial topological twist along the $\Sigma$ direction (see also \cite{Moore:1997dj,Nekrasov:2014xaa,Okuda:2013fea,Benini:2016hjo,Closset:2016arn,Kanno:2018qbn,Ueda:2019qhg, eckhard2020higher} for closely related analysis of similar systems and \cite{Closset:2019hyt} for a review).\footnote{Here $\beta$ is related to the equivariant parameter $t$ as $t=e^{-\beta}$. In the limit $\beta \to \infty$ , this QFT reduces to the usual Chern--Simons theory with compact gauge group. On the other hand, when $\beta \to 0$, its Hilbert space is identified with that of complex Chern--Simons theory.} 
This QFT has a one-form $Z(\tG)$ symmetry, inherited from that of the supersymmetric Chern--Simons--matter theory. 
This symmetry can in general have anomaly, which vanishes for a subgroup $Z\subset Z(\tG)$ when $k$ is a multiple of the smallest integer $k_{\tG/Z}$ satisfying \cite{Moore:1989yh,Dijkgraaf:1989pz}
\bea \label{eq:transgressive}
\frac{k_{\tG/Z}}{2} \langle \mu_i^\vee, \mu_i^\vee \rangle \in \bZ.
\eea
Here $\mu_i^\vee$ is a coweight such that $e^{2\pi i \mu_i^\vee}$ generates $Z$.\footnote{We recall that $Z(\tG)$ is identified with $\Lambda_w^\vee/\Lambda_R^\vee$ where $\Lambda_w^\vee$ and $\Lambda_R^\vee$ are coweight and coroot lattices of $\mathfrak{g}$ respectively. } As $\tG_k$ Chern--Simons theory is reduced on a circle, $\CT(\tG_k)$ inherits both a 0-form and a 1-form center symmetry. The anomaly in three dimensions manifests itself as a mixed anomaly between the two symmetries. In the rest of this section, we will discuss the symmetries of $\CT(\tG_k)$ and their gaugings. 

\subsection{0-form symmetry and its gauging} \label{sec:0formgauging}
The 0-form center symmetry of the equivariant Verlinde TQFT acts as a permutation of elements of $\tT^{\text{reg}}_{k,t}$.  In order to establish this statement we need to show that if $f$ is a solution satisfying $\chi_{k,t} (f) = e^{2\pi i \rho}$, then $gf$ is a solution for all $g \in Z(\tG)$. Equivalently, we need to show that $\chi_{k,t} (g) =1$. Furthermore, the fusion eigenvalue $\theta_t (f)^{-1/2}$ must be invariant under $f\to gf$. We will derive these results now. 

First, let us consider the case $t\to 0$. Here we want to show $\chi_k(g)=1$ and $\theta(gf) = \theta(f)$. We recall that $Z(\tG) \equiv \Lambda_w^\vee/\Lambda_R^\vee$ where $\Lambda_w^\vee$ and $\Lambda_R^\vee$ are coweight and coroot lattices respectively.  Following \cite{laredo1999positive}, we consider a representative $\mu_i^\vee$ of each $\Lambda_w^\vee/\Lambda_R^\vee$-coset defined to be the dual of each simple root $\alpha_i$ in the expansion $\vartheta = \sum_i m_i \alpha_i$ that appears with coefficient $1$.
Since any such $\alpha_i$ is a long root, the basic inner product identifies $\mu_1^\vee$ with a fundamental weight $w_i$, and we get
\bea 
\chi_k (g) = e^{2\pi i (k+h) w_i} = e^{2\pi i (k+h) w_i} 
\eea
which is the identity element in $\tT^\star \equiv \mathfrak{t}^\star/2\pi i \Lambda_w$.
This establishes the result $\chi_k (g)=1$. To show $\theta(gf) = \theta(f)$ we note that $e^{\alpha} (e^{2\pi i \mu_i^\vee}) = 1$ for all roots $\alpha$, and hence $\theta(gf) = \theta(f)$. This establishes the presence of $Z(\tG)^\zero$ symmetry in the $t\to 0$ limit.

As $t$ is deformed to a non-zero value, we need to show that the deformation of $\chi_k$ in equation \eqref{eq:deformedBethe} does not spoil the result. In particular, we want
$
\left( \frac{1-t e^{\alpha} (g) }{1-t e^{-\alpha} (g)} \right)^{\alpha} =1
$
for all  $\alpha \in \mathfrak{R}_+$.
This is straightforward as $e^{\alpha} (e^{2\pi i \mu_i^\vee}) = 1$, giving the required result $\chi_{k,t} (g)=1$. Finally, in order to show that $\theta_t(f) = \theta_t (gf)$ we need $\det (H_t^\dagger (gf)) = \det (H_t^\dagger (f)) $. In particular, using the expression for the Hessian of $D_t$ given in Lemma 7.7 of \cite{teleman2009index} we need that for each root $\alpha$,
\bea 
\frac{e^\alpha}{1-t e^\alpha}(gf) = \frac{e^\alpha}{1-t e^\alpha}(f)
\eea 
which is of course true again by the use of $e^{\alpha} (e^{2\pi i \mu_i^\vee}) = 1$.

 The orbits of $Z(\tG)$ action on $T^\reg_{k,t}/W$ are easy to compute in the $t\to 0$ limit, where $\tT^\reg_k/W$ is in bijection with $P_k$. $P_k$ admits an action of $Z(\tG)$ which is summarized in terms of Dynkin labels in \autoref{tab:2}.
As $t$ is deformed to a small non-zero value, the deformation of elements in $\tT^{\text{reg}}_{k}/W$ does not cross each other. Therefore, the orbits of the center action stay the same.

The partition function of $\CT(\tG_k)$ in the presence of a 0-form symmetry background $h \in H_1(\Sigma, Z)$ can be computed using equation \eqref{eq:Z_h},
\bea \label{eq:ZhinTGk}
\CZ_{h} = \sum\limits_{\lambda \in P_k^{\widetilde{Z}}} \theta_t(f_{\lambda,t})^{1-g}.
\eea
Here $\widetilde{Z}$ is the smallest subgroup of $Z$ such that $H_1(\Sigma, \widetilde{Z})$ contains $h$, and $P_k^{\widetilde{Z}}$ is the set of fixed points of $P_k$ under the $\widetilde{Z}$ action. Also, the partition function of $\CT(\tG_k)/Z^\zero$ TQFT is given through equation \eqref{eq:Z_main},\footnote{
Strictly speaking, we stated the results \eqref{eq:Z_main} and \eqref{eq:Z_h} in Section \ref{sec:FiniteGauging} for the case of cyclic groups only. Here when $\tG=Spin(4n,\C)$ and $Z=\Z_2 \times \Z_2$, $Z$ is not cyclic. However, the results \eqref{eq:ZhinTGk} and \eqref{eq:ZTGk/Z02} still hold. This is discussed in Appendix \ref{sec:product}.}
\bea \label{eq:ZTGk/Z02} 
\CZ_{\CT(\tG_k)/Z^\zero} = \sum\limits_{\lambda \in P_k/Z} |\Stab(\lambda)| \left( \frac{|Z| }{|\Stab(\lambda)|^2} \theta_t (f_{\lambda,t}) \right)^{1-g}.
\eea
 
\begin{table} \small%
\centering
\begin{tabular}{|p{4.0cm}|p{1.5cm}|p{9.75cm}|}
\hline
$\tG$ & $Z(\tG)$ & Action of $Z(\tG)$ generators  \\
\hline
$SL(n+1,\C)$, $n \geq 1$ & $\Z_{n+1}$ & 
$z( \lambda_1, \cdots, \lambda_{n-1}, \lambda_{n}) =
(\lambda_0, \lambda_1, \cdots, \lambda_{n-2}, \lambda_{n-1})$
\\
\hline
$Spin(2n+1,\C)$, $n \geq 2$ & $\Z_2$ & 
$z (\lambda_1,  \cdots, \lambda_{n-1}, \lambda_n) = 
( \lambda_0,  \lambda_2, \lambda_3, \cdots, \lambda_{n-1}, \lambda_n)$ \\
\hline

$Sp(n,\C)$, $n \geq 1$ & $\Z_2$ &
$z(\lambda_1, \cdots, \lambda_{n-1}, \lambda_{n}) =
(\lambda_{n-1}, \lambda_{n-2}, \cdots, \lambda_{1}, \lambda_{0})$
  \\
\hline

$Spin(4n,\C)$,  $n \geq 2$ & $\Z_2 \times \Z_2$ & 
$z_1 ( \lambda_1, \cdots, \lambda_{2n-1}, \lambda_{2n}) =
(\lambda_{0}, \lambda_{2}, \lambda_{3}, \cdots, \lambda_{2n-2}, \lambda_{2n}, \lambda_{2n-1})$
 \\
\cline{3-3}
                 &   & 
                 $z_2 ( \lambda_1, \cdots, \lambda_{2n-1}, \lambda_{2n}) =
(\lambda_{2n-1}, \lambda_{2n-2},  \cdots, \lambda_{1}, \lambda_{0})$ \\
\hline

$Spin(4n+2,\C)$, $n \geq 1$ & $\Z_4$ &
$z ( \lambda_1,  \cdots, \lambda_{2n}, \lambda_{2n+1}) =
(\lambda_{2n+1}, \lambda_{2n-1}, \lambda_{2n-2}, \cdots, \lambda_{1}, \lambda_{0})$
 \\     
 \hline
 
 $E_6$ & $\Z_3$ & 
 $z ( \lambda_1, \cdots, \lambda_{5}, \lambda_{6}) =
(\lambda_5, \lambda_4, \lambda_3, \lambda_{6}, \lambda_0, \lambda_{2})$ \\
 \hline
  $E_7$ & $\Z_2$ & 
   $z ( \lambda_1, \cdots, \lambda_{6}, \lambda_{7}) =
(\lambda_5, \lambda_4, \lambda_3, \lambda_{2}, \lambda_1, \lambda_0, \lambda_{7})$ \\
 \hline
\end{tabular}
\caption{\label{tab:2} The action of generators of $Z(G)$ on the Dynkin labels of weights in $P_k$. In each case, 
 $\lambda_0 = k- \langle \lambda, \vartheta \rangle$ where $\vartheta$ denotes the highest root of $G$.}
\end{table}

\subsubsection{Examples}

We will now present a few examples of $\CT(\tG_k)/Z^\zero$ theories. In each case, we will set $k \equiv 0 \pmod{k_{\tG/Z}}$ with the foresight of using these results in \autoref{sec:quantization}. 
As our first example, consider the simplest case $\tG=SL(2,\C)$, $Z=\Z_2$ and $k\equiv 0 \pmod 4$. The set $\tT^\reg_{k,t}/W$ is the set of solutions $\text{diag} (e^{i\varphi}, e^{-i\varphi})$ to the equation  
\bea 
e^{2i (k+2) \varphi} \left( \frac{1-t e^{2i \varphi}}{1-t e^{-2i \varphi}} \right)^2 = 1
\eea
with $\varphi \in (0,\pi)$. It has $k+1$ elements: $\varphi_j = \frac{(j+1) \pi}{k+2} + \mathcal{O}(t)$ for $j=0, 1, \cdots, k$. The equivariant Verlinde formula for $SL(2,\C)_k$ reads
\bea 
\CZ = \sum\limits_{j=0}^{k} \theta_t (e^{i\varphi_j})^{1-g} 
\eea
where
\bea 
\theta_t (e^{i\varphi}) = (1-t) \frac{4(\sin^2 \varphi) |1-t e^{2i\varphi}|^2 } {|\tT_k| \ \text{det} H_t^\dagger}
\eea
with 
\bea 
|\tT_k| \ \text{det} H_t^\dagger = 4 \left( \frac{k+2}{2} + \frac{2 t \cos 2\varphi - 2 t^2}{|1-t e^{2i\varphi}|^2} \right).
\eea

Under the identification $P_k \leftrightarrow \tT^{\reg}_k$, $j$ parameterizes the highest weights of an $\mathfrak{su}(2)$ representation. The action of $\Z_2$ on $P_k$ is $j \to k-j$; with the only fixed point being $j = \frac{k}{2}$. The corresponding element of $\tT^{\text{reg}}_{k,t}/W$ is
$\text{diag} \ (i,-i)$. We obtain that for any non-zero $h \in H_1 (\Sigma, \Z_2)$,
\bea \label{eq:ZhinSL2}
\CZ_{h} = \theta_t (i)^{1-g} = \left( \frac{\frac{k}{2}+1 +\left( \frac{k}{2}-1 \right) t}{ (1-t)(1+t)^3} \right)^{g-1}
\eea
and the partition function of $\CT(SL(2,\C)_k)/\Z_2^\zero$ is
\bea \label{eq:SL20gauged}
\CZ_{\CT(SL(2,\C)_k)/\Z_2^\zero} =  \sum\limits_{j=0}^{\frac{k}{2}-1} (2\theta_t (e^{i\varphi_j}))^{1-g} + 2 \left(\frac{\theta_t (i)}{2}\right)^{1-g}.
\eea

This example generalizes easily to the case of $G=SL(p,\C)$ for $p\geq 3$ prime, $Z=\Z_p$ and $k \equiv 0 \pmod p$. The only fixed point under the action action of $\Z_p$ on $P_k$ is $\frac{k}{p} \rho$; it exponentiates to the element $f_\star =   \text{diag} (1, \eta, \cdots, \eta^{p-1}) \in \tT^\text{reg}_{k,t}/W$ where $\eta = e^{2\pi i/p}$. For each non-zero $h \in H_1 (\Sigma, \Z_p)$,
\bea 
\CZ_h = \theta_t (f_\star)^{1-g}
\eea
and the partition function of $\CT(SL(p,\C)_k)/\Z_p^\zero$ is
\bea \label{eq:SLn0gauged}
\CZ_{\CT(SL(p,\C)_k)/\Z_p^\zero}=  \sum\limits_{\lambda \in (P_k\backslash \frac{k}{p}\rho)/\Z_p} (p\ \theta_t (f_{\lambda,t}))^{1-g} + p \left(\frac{\theta_t (f_\star)}{p}\right)^{1-g}.
\eea

As another example, we consider $\tG= SL(4,\C)$, $Z=\Z_4$ and $k \equiv 0 \pmod 8$. This example is different from the previous cases since $\Z_4$ has a non-trivial $\Z_2$ subgroup. The integrable weight $\frac{k}{4} \rho$ is fixed under the action of $\Z_4$; the corresponding element of $\tT^\reg_{k,t}/W$ is $f_\star = e^{i\pi/4} \ \text{diag} (1,i, -1,-i)$. The fixed point set $P_k^{\Z_2}$ is $\{(a, \frac{k}{2}-a, a) | a \leq \frac{k}{2}\}$. For each non-zero $h \in H_1 (\Sigma, \Z_2) \subset H_1 (\Sigma, \Z_4)$,
\bea \label{eq:SL4C_FZ2}
\CZ_h = \sum\limits_{\lambda \in F_{\Z_2}(P_k)} \theta_t (f_{\lambda,t})^{1-g}
\eea
and for $h \notin H_1 (\Sigma, \Z_2)$, 
\bea 
\CZ_h = \theta_t (f_\star)^{1-g} . 
\eea
The partition function of $\CT(SL(4,\C)_k)/\Z_4^\zero$ is 
\bea \label{eq:SL40gauged}
\CZ_{\CT(SL(4,\C)_k)/\Z_4^\zero} = \sum\limits_{(P_k \backslash P_k^{\Z_2})/\Z_4} (4 \theta_t (f_{\lambda,t}))^{1-g} 
+ 2 \sum\limits_{(P_k^{\Z_2} \backslash \{\frac{k}{4} \rho \})/\Z_2} \left(\theta_t (f_{\lambda,t})\right)^{1-g} 
+ 4 \left(\frac{\theta_t (f_\star)}{4}\right)^{1-g}. 
\eea

Finally, we consider $\tG=SL(4,\C)$, $Z=\Z_2 \subset \Z_4$ and $k\equiv 0 \pmod 2$. Here, $\CZ_h$ is the same as in \eqref{eq:SL4C_FZ2} for all non-zero $h \in H_1(\Sigma, \Z_2)$, and the partition function of $\CT(SL(4,\C)_k)/\Z_2^\zero$ is
\bea 
\CZ_{\CT(SL(4,\C)_k)/\Z_2^\zero} = \sum\limits_{(P_k \backslash P_k^{\Z_2})/\Z_2} (2 \theta_t (f_{\lambda,t}))^{1-g} + 2 \sum\limits_{ P_k^{\Z_2} } \left(\frac{ \theta_t (f_{\lambda,t})}{2}\right)^{1-g}.
\eea

\subsection{1-form symmetry and its gauging} \label{sec:1formgauging}
The 1-form center symmetry of $\CT(\tG_k)$ acts on the elements of $\tT^\reg_{k,t}/W$ by  multiplication with roots of unity. This action can be described as follows. Let $\lambda \in P_k$ be a highest integrable weight of $\tG$; the center $Z(\tG)$ acts on the associated $\tG$-representation as $e^{2\pi i \mu_i^\vee} \to e^{2\pi i \lambda(\mu_i^\vee)} \mathbb{1}$. We assign this phase to $f_{\lambda,t}$; thus giving the required $Z(\tG)$ action.\footnote{This is consistent with the familiar action of one-form symmetry in the anyon basis for $t=0$. There, the basis elements are labelled by level $k$ integrable representations and the center symmetry acts by the action of $Z(\tG)$ on the Weyl alcove. In particular, for $z \in Z(\tG)$ and $\lambda,\chi \in P_k$, $N_{z\lambda}{}^\chi = \delta_{z(\lambda)}{}^\chi$. The basis transformation is carried out by the S-matrix and we have
\bea \nonumber
\sum\limits_{\rho, \sigma} S_{\lambda}{}^{ \rho} e^{2\pi i \rho(\mu_i^\vee)} \delta_{\rho}{}^{ \sigma} (S^{-1})_{\sigma}{}^{ \chi}
= \sum\limits_{\rho} S_{\lambda}{}^{ \rho} e^{2\pi i \rho(\mu_i^\vee)} (S^{-1})_{\rho}{}^{ \chi}
= \sum\limits_{\rho} S_{z(\lambda)}{}^{ \rho} (S^{-1})_{\rho}{}^{ \chi}
= \delta_{z(\lambda)}{}^{ \chi}
\eea
Here, in the second equality we have used the result $S_{z(\lambda)}{}^{ \rho} = S_{\lambda}{}^{ \rho} e^{2\pi i \rho(\mu_i^\vee)}$. See section 14.6.4 of \cite{DiFrancesco:1997nk} for a derivation.
}

In the presence of an $e^{2\pi i \mu_i^\vee} \in Z$ background, the partition function of $\CT(\tG_k)$ is (see equation \eqref{eq:1form_w_background}), 
\bea \label{eq:1-form_background_G/Z}
\CZ_{\mu_i^\vee} = \sum\limits_{\lambda \in P_k} e^{2\pi i  \lambda (\mu_i^\vee)} \theta_t (f_{\lambda,t})^{1-g}.
\eea
Gauging the symmetry projects out weights with $e^{2\pi i \lambda(\mu_1^\vee)} \neq 1$ (for $e^{2\pi i \mu_1^\vee}$ a generator of $Z$). The partition function of $\CT(\tG_k)/Z^\one$ is (see equation \eqref{eq:1formgauged}),
\bea \label{eq:T/Zgauged}
\CZ_{\CT(\tG_k)/Z^\one} = \sum\limits_{\lambda \in P_k(G)} \left(|Z| \theta_t (f_{\lambda,t})\right)^{1-g}.
\eea
Here, $P_k (G) \subset P_k$ is the subset of highest weights that label representations of $G = \tG/Z$. 

\subsubsection{Examples}

Now we will study some examples. Consider the case $\tG= SL(2,\C)$ and $Z=\Z_2$. Here $\mu_1^\vee = \frac{1}{2} \ \text{diag}(1,-1)$, and in the notation of Section \ref{sec:0formgauging},
\bea 
\CZ_{\mu_1^\vee} = \sum\limits_{j=0}^k (-1)^j \theta_t (e^{i\varphi_j})^{1-g}.
\eea
The partition function of $\CT(SL(2,\C)_k)/\Z_2^\one$ gets contribution from even weights only, i.e.
\bea 
\CZ_{\CT(SL(2,\C)_k)/\Z_2^\one} = \sum\limits_{\substack{j=0 \\ j\  \text{even}}}^{k} \left(2 \theta_t (e^{i\varphi_j}) \right)^{1-g}.
\eea

This example generalizes easily to the case of $SL(n,\C)$ and $Z=\Z_m$. Let $n_t(\lambda) = \sum\limits_{j=1}^{n-1} j \lambda_j$ be the $n$-ality of the representation with highest weight $\lambda$, then 
\bea 
\CZ_{\mu_1^\vee} = \sum\limits_{\lambda \in P_k} e^{2\pi i n_t(\lambda)/m} \theta_t (f_{\lambda,t})^{1-g}.
\eea
Summing over $\Z_m$ backgrounds gives 
\bea 
\CZ_{\CT(SL(n,\C)_k)/\Z_m^\one}  = \frac{1}{m} \sum\limits_{\lambda \in P_k} \left(\sum\limits_{s=1}^m e^{2\pi i s n_t(\lambda)/m} \right) \left( m \theta_t (f_{\lambda, t}) \right)^{1-g}.
\eea
The sum inside the brackets equals $m$ when $n_t (\lambda) = 0 \mod m$, and zero otherwise. That is, it is non-zero precisely when $\lambda$ gives a representation of $SL(n,\C)/\Z_m$. We obtain
\bea 
\CZ_{\CT(SL(n,\C)_k)/\Z_m^\one} = \sum\limits_{\lambda \in P_k(SL(n,\C)/\Z_m)} \left(m\theta_t (f_{\lambda, t})\right)^{1-g}.
\eea

\subsection{Quantization of $\mathcal{M} (G)$} \label{sec:quantization}
In this section, we will use the results from the previous two sections to compute the Hitchin character for the moduli space $\cM (G)$ of semistable $G$-Higgs bundles.
This moduli space is disconnected with $\pi_0 (\cM(G)) = \pi_1(G)= Z$; the second Stiefel--Whitney class $w_2$ of the $G$-Higgs bundle on the surface parameterizes the various components.

For a fixed $w_2$, $\cM_{w_2}(G)$ is the quotient $\cM_d/\mathcal{G}$ of the moduli space $\cM_d$ of semistable $\tG$-Higgs bundles of degree $d=w_2$ by $\mathcal{G} = H_1 (\Sigma, Z)$. The action of $\mathcal{G}$ on $\cM_d$ lifts naturally to an action on the pre-quantum line bundle $\cL^k$. This gives the space of sections $H^0(\cM_d, \cL^k)$ the structure of a $\CG$-module. Classically, the moduli space $\cM(G)$ is the phase space of $G$ Chern--Simons theory on $S^1 \times \Sigma$. As this theory is well-defined only for $k\equiv 0 \pmod{ k_{\tG/Z}}$ (see equation \eqref{eq:transgressive}), we will consider pre-quantum line bundles on $\cM_{w_2}$ whose pullback to $\cM_d$ is $\cL^k$ for these values. (By an abuse of notation we will also denote the corresponding line bundle on $\cM_{w_2}$ as $\cL^k$.) The space of sections $H^0 (\cM_{w_2} (G), \cL^k)$ is isomorphic to the subspace of $\CG$-invariant sections of $H^0 (\cM_d, \cL^{k})$.

The actions of $\CG$ and $\C^\star$ on $\cL^k$ commute, therefore, the moduli space  $\cM_{w_2}(G)$ also admits a $\C^\star$ action.
\bea 
H^0 (\cM_{w_2} (G), \cL^k) = \oplus_{n} H^0_n (\cM_{w_2} (G), \cL^k).
\eea
We define the Hitchin character for $\cM_{w_2}(G)$, 
\bea \label{eq:G/Zcharacter}
\text{dim}_t \ H^0 (\cM_{w_2} (G) , \cL^k) := \sum\limits_{n=0}^\infty t^n \ \text{dim} H^0_n (\cM_{w_2} (G),\cL^k).
\eea
As the space $H^0(\cM_{w_2}(G),\cL^k)$ is isomorphic to the $\CG$-invariant subspace of $H^0 (\cM_d,\cL^k)$, we have the equality
\bea 
\text{dim} H^0_n (\cM_{w_2} (G),\cL^k) =  \frac{1}{|\CG|} \sum\limits_{h \in \CG} \Tr (h| H^0_n (\cM_{d=w_2} (\tG), \cL^{k})).
\eea
Let the formal series containing all the traces for a fixed $h$ be 
\bea \label{eq:eq_trace}
 \Trt (h| H^0 (\cM_d  , \cL^{k})) = \sum\limits_{n=0}^\infty \Tr (h| H^0_n (\cM_d, \cL^{k})) t^n.
 \eea
 We will call this trace the \textit{equivariant trace}. The Hitchin character is just the average of the equivariant traces. 
\bea \label{eq:characterandtrace}
\text{dim}_t \ H^0 (\cM_{w_2} (G) , \cL^k) = \frac{1}{|\CG|} \sum\limits_{h \in \CG} \Trt (h| H^0 (\cM_{d=w_2}  , \cL^{k})).
\eea

In the rest of this section, we will compute the equivariant trace as well as the Hitchin character for $\cM_{w_2} (G)$ using methods in 2d TQFTs.   Our main observation is that the equivariant trace \eqref{eq:eq_trace} is equal to the partition function of $\CT(\tG_k)$ in the presence of the 0-form symmetry background $h\in H_1(\Sigma,Z)$ and the 1-form symmetry background $d\in Z$. For $d=0$, this trace is equal to $\CZ_h$ in equation \eqref{eq:ZhinTGk}. We will give an argument for this identification now.

We recall that the space of sections $H^0 (\cM_{d=0} , \cL^{k})$ is identified with the Hilbert space of complex $\tG$ Chern--Simons theory (or alternatively, the $\beta$-deformed Chern--Simons theory) on the surface $\Sigma$ \cite{Gukov:2015sna}. 
$\tG$ Chern--Simons theory has a 1-form center symmetry with an invertible topological line operator labelled by each $d\in Z$. 
The insertion of this line along a temporal cycle amounts to creating the defect Hilbert space $H^0 (\cM_{d} , \cL^{k})$,\footnote{The insertion of the symmetry generating line along $S^1$ changes the classical phase space to $\cM_{d}$. Its quantization $H^0 (\cM_{d} , \cL^{k})$ gives the defect Hilbert space.}
while its insertion along a spatial cycle creates an action of the line operator on the Hilbert space. 
The partition function $\CZ_{CS}(S^1 \times \Sigma)$ computes the trace $\Trt (h|H^0(\cM_d, \cL^k))$ when these line operators are inserted.

On the other hand, the insertion of these line operators along the spatial (temporal) cycles amounts to creating a 0-form (1-form) symmetry background of $\CT(\tG_k)$. 
 $\CZ_{CS}(S^1\times \Sigma)$ now computes the partition function of $\CT(\tG_k)$ in the presence of these backgrounds, thus establishing the afore-mentioned identification.

We will now compute the trace $\Trt (h|H^0(\cM_d, \cL^k))$ and the Hitchin character $ \text{dim}_t \ H^0 (\cM_{w_2} (G) , \cL^k) $ using 2d TQFT methods. These computations will build on the computations presented in Section \ref{sec:0formgauging} and Section \ref{sec:1formgauging}.

First, consider the case $d=0$. In this case, $\CZ_h = \Trt (h|H^0(\cM_{d=0}, \cL^k))$ was already computed in equation \eqref{eq:ZhinTGk}. The Hitchin character can be computed as follows: it is $\frac{1}{|Z|^{g}}$ times the partition function of $\CT(\tG_k)/Z^\zero$ TQFT \eqref{eq:ZTGk/Z02},\footnote{
The extra factor $\frac{1}{|Z|^{g}}$ appears because of the difference in normalizations of $\mathcal{Z}_{\CT(\tG_k)/Z^\zero}$ and $\text{dim}_t \ H^0 (\cM_{w_2=0} (G) , \cL^k)$. In particular, 
\bea \nonumber
\mathcal{Z}_{\CT(\tG_k)/Z^\zero} = \frac{1}{|Z|^{g}} \sum\limits_{h \in H_1(\Sigma, Z)} \CZ_h
\eea
 while 
\bea \nonumber
\text{dim}_t \ H^0 (\cM_{w_2=0} (G) , \cL^k) = \frac{1}{|Z|^{2g}} \sum\limits_{h \in H_1(\Sigma, Z)} \CZ_h.
\eea 
The Hitchin character for $\cM_{w_2} (G)$ is \textit{not} a partition function of a 2d TQFT. This is expected as cutting and gluing do not preserve the second Stiefel--Whiney class of a $G$-Higgs bundle.}
\bea \label{eq:HitchinchG/Zw20}
\text{dim}_t \ H^0 (\cM_{w_2=0} (G) , \cL^k) = \frac{1}{|Z|^{g}} 
\sum\limits_{\lambda \in P_k/Z} |\Stab(\lambda)| \left( \frac{|Z|}{|\Stab(\lambda)|^2} \theta_t (f_{\lambda,t}) \right)^{1-g}.
\eea

For $d\neq 0$, $\text{dim}_t H^0 (\cM_d, \cL^k)$ is the  partition function of $\CT(\tG_k)$ in the presence of an $e^{2\pi i \mu_i^\vee} \in Z$ 
background \eqref{eq:1-form_background_G/Z}.\footnote{
We remark that our notation for elements $\{e^{2\pi i \mu_i^\vee}\}$ of $Z$ is multiplicative, while that for degree $d$ is additive. 
} 
The equivariant trace $\Trt  (h|H^0 (\cM_d, \cL^k))$ can be computed by further introducing a $Z^\zero$ background.
This restricts the sum in \eqref{eq:1-form_background_G/Z} to $\widetilde{Z}$ invariant operators where $\widetilde{Z}$ is the smallest subgroup of $Z$ such that $H_1(\Sigma, \widetilde{Z}) \subset H_1 (\Sigma, Z)$ contains $h$,
\bea \label{eq:traced}
\Trt  (h|H^0 (\cM_d, \cL^k)) = \sum\limits_{\lambda \in P_k^{\widetilde{Z}}} e^{2\pi i \lambda (\mu_i^\vee)} \theta_t (f_{\lambda, t})^{1-g}.
\eea
The Hitchin character $\text{dim}_t \ H^0 (\cM_{w_2} (G) , \cL^k) $ is obtained by averaging the equivariant traces over all $h \in H_1 (\Sigma,Z)$. In the absence of a mixed anomaly, i.e. when $k \equiv 0 \pmod{k_{\tG/Z}}$, the actions of $Z^\zero$ and $Z^\one$ on $T^\reg_{k,t}/W$ commute. This implies that each element of a $Z^\zero$ orbit has the same phase $e^{2\pi i \lambda (\mu_i^\vee) }$ under $Z^\one$ action. Furthermore, a computation similar to that of Section \ref{sec:gauging} gives
\bea \nonumber 
\text{dim}_t \ H^0 (\cM_{w_2=d} (G) , \cL^k) = \frac{1}{|Z|^{g}} \sum\limits_{\lambda \in P_k/Z}  e^{2\pi i \lambda (\mu_i^\vee)}  |\Stab(\lambda)| \left( \frac{|Z|}{|\Stab(\lambda)|^2} \theta_t (f_{\lambda, t}) \right)^{1-g}. \\ \label{eq:dcharacter}
\eea
In the $t\to 0$ limit, the fact that dimension of sections of $\CL^k$ over the individual component labeled by $d$ takes such a nice form was observed by Beauville \cite{beauville1996verlinde}. This phenomenon, while not having a simple CFT interpretation, is very natural from the 2d TQFT perspective, as the 1-form and 0-form symmetry, although they share the same 3d origin, are independent symmetries of the theory. 

Summing over $w_2$ gives the partition function of $\beta$-deformed $G$ complex Chern--Simons theory on $S^1 \times \Sigma$. We define $\CT(G_k)$ to be the circle reduction of this QFT, then 
\bea 
\mathcal{Z}_{\CT(G_k)} = \sum\limits_{\lambda \in P_k(G)/Z} |\Stab (\lambda)| \left( \frac{|Z|^2}{|\Stab(\lambda)|^2} \theta_t (f_{\lambda, t})
\right)^{1-g}.
\eea
Here $P_k(G)/Z$ is the set of all orbits of $P_k(G)$ under the action of $Z$.

\subsubsection{Examples}

The examples in this section will build on the examples studied in Section \ref{sec:0formgauging} and Section \ref{sec:1formgauging}. Consider first $G=SL(2,\C)$, $Z=\Z_2$ and $k \equiv 0 \pmod 4$. The equivariant trace was computed in Equation \eqref{eq:ZhinSL2},
\bea 
\Trt  (h|H^0 (\cM_d, \cL^k)) = \theta_t(i)^{1-g}.
\eea
It is independent of $d$ as $e^{2\pi i \frac{k}{2}\rho (\mu_i^\vee)} = 1$ (cf. equation \eqref{eq:traced}). The Hitchin character of $\cM_{w_2}(PSL(2,\C))$ for $w_2=0,1$ is 
\bea
\text{dim}_t H^0 (\cM_{w_2}(PSL(2,\C), \cL^k) &=& \frac{1}{2} \sum\limits_{j=0}^{\frac{k}{2}-1} (-1)^{j  w_2} (4 \theta_t (e^{i\varphi_j}))^{1-g} + \theta_t (i)^{1-g}.
\eea
Summing over $w_2$ gives the partition function of $\CT(PSL(2,\C)_k)$,
\footnote{
The sum in $Z_{\CT(PSL(2,\C)_k)}$ is over the set of solutions to the ``$PSL(2,\C)$ Bethe equation", i.e. 
$e^{i(k+2) \varphi} \left(\frac{1-t e^{-2i\varphi}}{1-t e^{2i\varphi}} \right) = -1$. This equation was derived in \cite{Gukov:2015sna} by studying $U(2)$ Bethe equations and appropriately decoupling some degrees of freedom. We note that the range of $\varphi$ is $\left(0, \frac{\pi}{2} \right]$ here; it is halved compared to the case of $SL(2,\C)$ where $\varphi \in (0,\pi)$ due to the quotient $SL(2,\C)/\Z_2 \equiv PSL(2,\C)$. The fixed point of $\Z_2$ i.e. $\varphi = \frac{\pi}{2}$ contributes two operators to the diagonal basis of $\CT(PSL(2,\C)_k)$.}
\bea 
Z_{\CT(PSL(2,\C)_k)} = \sum\limits_{\substack{j=0 \\ j \ \text{even}}}^{\frac{k}{2}-2} (4\theta_t (e^{i\varphi_j}))^{1-g} + 2 (\theta_t(i))^{1-g} .
\eea

This example generalizes easily to the case of $G=SL(p,\C)$, $Z=\Z_p$ for $p \geq 3$ prime and $k\equiv 0 \pmod p$. Since the only one fixed point $\frac{k}{p}\rho$ has $n$-ality equal to 0 mod $p$, we have
\bea 
\Trt  (h|H^0 (\cM_d, \cL^k)) = \theta_t(f_\star)^{1-g}
\eea
where $f_\star =   \text{diag} (1, \eta, \cdots, \eta^{n-1})$ and $\eta = e^{2\pi i/p}$.
The Hitchin character for $\cM_{w_2}(PSL(p,\C))$ is 
\bea \nonumber
\text{dim}_t H^0 (\cM_{w_2}(PSL(p,\C), \cL^k) = \frac{1}{p}  \sum\limits_{\lambda \in (P_k\backslash \frac{k}{p}\rho)/\Z_p} e^{2\pi i w_2 n_t (\lambda)/p} (p^2 \theta_t (f_{\lambda,t}))^{1-g} + \theta_t (f_\star)^{1-g} \\ 
\eea
where $w_2 = 0, 1, \cdots, p-1$. Summing over $w_2$ gives the partition function of $\CT(PSL(p,\C)_k)$.
\bea 
Z_{\CT(PSL(p,\C)_k)} = \sum\limits_{\lambda \in (P_k (PSL(p,\C)) \backslash \frac{k}{p}\rho)/\Z_p} \left(p^2 \theta_t (f_{\lambda,t}) \right)^{1-g} + p \ \theta_t (f_\star)^{1-g}.
\eea

Finally, we consider the case of $G=PSL(4,\C)$ for $k \equiv 0 \pmod 8$. For non-zero $h \in H_1(\Sigma, \Z_2)$, $\CZ_h$ gets contribution from $P_k^{\Z_2} = \{(a, \frac{k}{2}-a, a) | a \leq \frac{k}{2}\}$; $n$-ality of the element $(a, \frac{k}{2}-a, a)$ is $k+2a$. This gives
\bea 
\Trt  (h|H^0 (\cM_d, \cL^k)) =
 \sum\limits_{\substack{\lambda = (a, \frac{k}{2}-a, a) | \\ a\leq \frac{k}{2}} } 
 (-1)^{ad} \theta_t (f_{\lambda, t})
\eea
where $d=0,1,2,3$. For $h \notin H_1(\Sigma, \Z_2)$, 
\bea 
\Trt  (h|H^0 (\cM_d, \cL^k)) =\theta_t (f_\star)
\eea
where $f_\star = e^{i\pi/4} \ \text{diag} (1,i, -1,-i)$. The Hitchin characters $ \text{dim}_t H^0 (\cM_{w_2}(PSL(n,\C), \cL^k))$ for $w_2 = 0,1,2,3$ are 
\bea \nonumber
\frac{1}{4}  \sum\limits_{(P_k \backslash P_k^{\Z_2})/\Z_4}
e^{2\pi i w_2 n_t (\lambda)/4}
 \left( 16 \theta_t (f_{\lambda,t})\right)^{1-g} 
+ \frac{1}{2} \sum\limits_{\substack{\lambda = (a, \frac{k}{2}-a, a) | \\ a \leq \frac{k}{2}, a \neq \frac{k}{4}} }
(-1)^{a w_2}
 \left(4 \theta_t (f_{\lambda,t}) \right)^{1-g} 
+  \theta_t (f_\star)^{1-g} .
\eea
 Finally, the partition function of $\CT(PSL(4,\C)_k)$ is
\bea \nonumber
\CZ_{\CT(PSL(4,\C)_k)} = 
\sum\limits_{
\substack{\lambda \in P_k(PSL(4,\C))/\Z_4\\ |\lambda|=4}} 
 \left(16 \theta_t (f_{\lambda,t}) \right)^{1-g} 
 + 2 \sum\limits_{
\substack{\lambda \in P_k(PSL(4,\C))/\Z_4\\ |\lambda|=2}} 
  (4 \theta_t (f_{\lambda, t}))^{1-g} 
+ \theta_t (f_\star)^{1-g}.
\eea
Here $\lambda$ denotes a $\Z_4$-orbit of $P_k(PSL(4,\C))$ and $|\lambda|$ denotes the length of this orbit.

\section{Equivariant Verlinde algebra for $PSL(2,\mathbb{C})$}  \label{sec:PSL2Calgebra}

In this section, we will study the equivariant Verlinde algebra for $PSL(2,\C)$ Chern--Simons theory. In the case of $SL(2,\C)$ Chern--Simons, the equivariant Verlinde algebra was studied in \cite{Gukov:2015sna}. It was extended to the case of $SL(3,\C)$ in \cite{Gukov:2016lki}, and to arbitrary simple and simply-connected complex Lie groups in \cite{Andersen:2016hoj}. These algebras have the geometric interpretation of Hitchin characters for moduli space of parabolic semi-stable $G$-Higgs bundles. 

Here, we extend their results to non-simply connected groups. We will focus mostly on the case of $PSL(2,\C)$ as a concrete example, obtaining our results by gauging the $\Z_2^\zero \times \Z_2^\one$ symmetry of $\CT(SL(2,\C)_k)$ in its parabolic basis. Our methods can be easily generalized to study the equivariant Verlinde algebra for arbitrary simple complex Lie groups $G$, however this generalization will be studied elsewhere. Towards the end of this section, we will remark on the relation of the $PSL(2,\C)$ algebra to the quantization of the moduli space of parabolic semi-stable $PSL(2,\C)$-Higgs bundles.

Let us first recall the details of the equivariant Verlinde algebra for $SL(2,\C)$. The basis elements in parabolic basis are labelled by the integrable highest weights of $\mathfrak{su}(2)$ at level $k$: $w^0, w^1, \cdots, w^k$. The fusion rules of these operators are
\bea \label{eq:SL2fusion}
f^{a_1 a_2 a_3} = 
\begin{cases} 
1 &\mbox{if }a_1 + a_2 + a_3 \in 2\mathbb{Z} \ \text{and} \Delta \leq 0 \\
t^{\Delta /2} &\mbox{if } a_1 + a_2 + a_3 \in 2\mathbb{Z} \ \text{and} \Delta  > 0 \\
0 & \mbox{otherwise} 
\end{cases}
\eea
where $\Delta = \text{max} (d_0, d_1, d_2, d_3)$ and 
\bea \nonumber
d_0 &=& a_1 + a_2 + a_3 - 2k \\ \nonumber
d_1 &=& a_1 - a_2 - a_3 \\ \nonumber
d_2 &=& a_2 -  a_3 - a_1 \\ \nonumber
d_3 &=& a_3 - a_1 - a_2 . 
\eea
 The identity element of the algebra is $w^0 - tw^2$ and the metric is given by
\bea \label{eq:SL2metric}
\eta_{a_1 a_2} = \text{diag} \left(\frac{1}{1-t^2}, \frac{1}{1-t}, \cdots, \frac{1}{1-t}, \frac{1}{1-t^2}\right). 
\eea
The algebra has a $\Z_2^\one$ symmetry generated by $w^k - t w^{k-2}$. In particular, this operator fuses with itself to the identity element,
\bea 
(w^k - t w^{k-2}) \times (w^k - tw^{k-2}) = w^0 - tw^2
\eea
and it acts on local operators as 
\bea 
(w^k - tw^{k-2}) \times w^j = w^{k-j}. 
\eea

The zero form center symmetry $\Z_2^\zero$ is generated by a line operator of $\CT(SL(2,\C)_k)$.  As $\CT(SL(2,\C)_k)$ is circle reduction of a three-dimensional QFT (the $\beta$-deformed $SL(2,\C)$ Chern--Simons theory), the 0-form symmetry generating line of $\CT(SL(2,\C)_k)$ is identified with the 1-form symmetry generating line of the $\beta$-deformed $SL(2,\C)$ Chern--Simons theory. In the $t\to 0$ limit (i.e. the case of compact group $SU(2)$ Chern--Simons theory), the action of this line operator is well known. It acts via braiding as $w^k w^{j} = (-1)^j w^j$. The phase is restricted to be a square root of 1, therefore, it does not change when $t$ is deformed. We get
\bea 
(w^k - t w^{k-2}) \ w^j = (-1)^j w^j . 
\eea

We will now discuss gauging these symmetries. In three dimensions, gauging 1-form symmetry of $SL(2,\C)$ Chern--Simons theory gives $PSL(2,\C)$ Chern--Simons theory. In two dimensions, gauging $\Z_2^\zero \times \Z_2^\one$ of $\CT(SL(2,\C)_k)$ will give the equivariant Verlinde TQFT $\CT(PSL(2,\C)_k)$ that encodes fusion rules of the $\beta$-deformed $PSL(2,\C)_k$ Chern--Simons theory. We will restrict to the case $k\equiv 0 \pmod 4$ so that the mixed anomaly between $\Z_2^\zero$ and $\Z_2^\one$ vanishes.

Now consider first gauging the $\Z_2^\one$ symmetry.\footnote{The final result, i.e. the set of fusion coefficients of $\CT(PSL(2,\C)_k)$ is independent of the order of gauging.}
 The gauge invariant operators are
\bea \label{eq:wj}
x^j = \frac{1}{2} (w^j + w^{k-j} )
\eea
for $j = 0, 1, \cdots, \frac{k}{2}$ where the normalization factor is fixed by comparison with the partition function of $\CT/\Z_2^\one$ gauged theory given in Section \ref{sec:gauging}.
The fusion rules of $\CT(SL(2,\C)_k)/\Z_2^\one$ are easy to derive. The derivation is postponed until Appendix \ref{sec:Z21derivation}; here we give the final result,
\bea \label{eq:fprimeabc1}
(f')^{ab}{}_{c} &=&  f^{ab}{}_c + f^{(k-a)b}{}_c \quad ; \quad c \neq \frac{k}{2} \\ \label{eq:fprimeabc2}
(f')^{ab}{}_{\frac{k}{2}} &=& f^{ab}{}_{\frac{k}{2}}. 
\eea

$\CT(SL(2,\C)_k)/\Z_2^\one$ inherits a $\Z_2^\zero$ symmetry from its parent TQFT. This symmetry acts as $x^j \to (-1)^j x^j$. Gauging $\Z_2^\zero$ projects out $x^j$ for odd $j$, while $x^j$ with even $j$ survive.
 $x^{\frac{k}{2}}$ plays a distinguished role under gauging; it splits into two operators $x^{{\frac{k}{2}}^\one}$ and $x^{{\frac{k}{2}}^{(2)}}$,\footnote{This fact is familiar from the point of view of compact group Chern--Simons theory \cite{Moore:1989yh,Delmastro:2021xox}. It is also consistent with the results obtained for $\CT(PSL(2,\C)_k)$ in Section \ref{sec:0formgauging}. In the diagonal basis $ T^\reg_{k,t}/W$, $\Z_2^\zero$ has a unique fixed point, $\text{diag} (i,-i)$. The gauged theory $\CT(SL(2,\C)_k)/\Z_2^\zero$ inherits two copies of this operator. The claim here is that in the parabolic basis, the operator that gives two copies is $x^{\frac{k}{2}}$. An understanding of why it is the $x^{\frac{k}{2}}$ operator (and not $x^j$ for any other value of $j$) that splits purely from a 2D TQFT point of view would be very interesting.}
 \bea \label{eq:x+}
 x^{\frac{k}{2}} = x^{{\frac{k}{2}}^\one} + x^{{\frac{k}{2}}^\two}.
 \eea
 The dual $\Z_2^\zero$ symmetry of $\CT(PSL(2,\C)_k)$ exchanges $x^{{\frac{k}{2}}^\one} $ and $x^{{\frac{k}{2}}^\two}$.
Overall, the operators of $\CT(PSL(2,\C)_k)$ are 
\bea \label{eq:basis_elements}
x^0 , x^2, \cdots, x^{\frac{k}{2}-2}, x^{{\frac{k}{2}}^\one}, x^{{\frac{k}{2}}^{(2)}}.
\eea

It is convenient to split the fusion rules of these operators into six kinds depending on how the $x^{\frac{k}{2}^{(i)}}$ operators appear. Let $a,b,c$ denote the highest weights $0, 2, \cdots, \frac{k}{2}-2$, and $\mu,\nu, \rho$ denote $\frac{k}{2}^{(i)}$, then the six different kinds are
\bea \label{eq:psl2cfsec4}
\widetilde{f}^{ab}{}_c, \quad \widetilde{f}^{ab}{}_\mu, \quad \widetilde{f}^{a\mu}{}_b, \quad \widetilde{f}^{a\mu}{}_\nu, \quad \widetilde{f}^{\mu \nu}{}_a, \quad \widetilde{f}^{\mu \nu}{}_\rho.
\eea
Using equation \eqref{eq:x+} and the dual $\Z_2^\zero$ symmetry, the first three sets of these are easily fixed. We postpone the derivation until Appendix \ref{sec:Z20derivation}; here we give the final result.
\bea \label{eq:fabc}
\widetilde{f}^{ab}{}_c &=& f^{ab}{}_c + f^{(k-a)b}{}_c \\ \label{eq:fabmu}
\widetilde{f}^{ab}{}_\mu &=& f^{ab}{}_{\frac{k}{2}} \\ \label{eq:famub}
\widetilde{f}^{a\mu}{}_b &=& f^{a\frac{k}{2}}{}_{b}.
\eea

The last three sets in \eqref{eq:psl2cfsec4} require a little more work. In addition to the $\Z_2^\zero$ dual symmetry, here we also need to use associativity of the TQFT and a positivity condition on the fusion coefficients. This condition is described as follows. A fusion coefficient of $\CT(PSL(2,\C)_k)$ is the $\C^\star$ character of the Hilbert space of $\beta$-deformed $PSL(2,\C)$ Chern--Simons theory on $S^1 \times S^2$ with three line operators inserted at three points of $S^2$. The character $\sum\limits_{n=0}^\infty C_n t^n$ must have positive integer coefficients $C_n \in \Z_{\geq 0}$. After applying these constraints, we find that 
\bea \label{eq:famunu}
\widetilde{f}^{a\mu}{}_\nu = 
\begin{cases} 
\frac{1}{1-t^2} &\mbox{if } a \equiv 0 \pmod{4} \ \& \ \mu = \nu, \ \text{or} \ a \equiv 2 \pmod{4} \ \& \ \mu \neq \nu \\
\frac{t}{1-t^2} &\mbox{if }a \equiv 2 \pmod{4} \ \& \ \mu = \nu, \ \text{or} \ a \equiv 0 \pmod{4} \ \& \ \mu \neq \nu  ,
\end{cases}
\eea
\bea \label{eq:fmunua}
\widetilde{f}^{\mu \nu}{}_a = 
\begin{cases} 
0 &\mbox{if } a \equiv \frac{k}{2}+2 \pmod{4} \ \& \ \mu = \nu, \ \text{or} \ a \equiv \frac{k}{2} \pmod{4} \ \& \ \mu \neq \nu \\
\eta_{aa}  &\mbox{if } a \equiv \frac{k}{2}+2 \pmod{4} \ \& \ \mu \neq \nu, \ \text{or} \  a \equiv \frac{k}{2} \pmod{4} \ \& \ \mu = \nu
\end{cases}
\eea
and 
\bea \label{eq:fmunurho1}
\widetilde{f}^{\mu \nu}{}_\rho =
\begin{cases} 
0 &\mbox{if } \mu = \rho \neq \nu \\
\frac{t^l}{1-t^2} &\mbox{if } \mu = \rho = \nu \\
\frac{t^{1-l}}{1-t^2}  &\mbox{if } \mu = \nu \neq \rho.
\end{cases}
\eea
In \eqref{eq:fmunua}, $\eta_{aa}$ are components of $SL(2,\C)$ metric \eqref{eq:SL2metric}, and in \eqref{eq:fmunurho1}, $l$ is defined to be a new variable: $l = \frac{k}{4} \bmod 2$. These fusion coefficients match with the fusion coefficients of $SO(3)_k$ Chern--Simons theory previously found in the literature \cite{Brustein:1988vb,Delmastro:2021xox}.

The identity element of $\CT(PSL(2,\C)_k)$ is $x^0 - t x^2$, thus giving the following block diagonal form of the metric,
\bea 
\widetilde{\eta}_{ab} = \text{diag} \left(\frac{1}{1-t^2}, \frac{1}{1-t}, \cdots, \frac{1}{1-t} \right)
\eea
and 
\bea 
\widetilde{\eta}_{\mu \nu} = \frac{1}{1-t^2} \times
\begin{cases} 
t^l &\mbox{if } \mu = \nu  \\
t^{1-l} &\mbox{if } \mu \neq \nu
\end{cases}
\eea

We propose that the fusion coefficients \eqref{eq:fabc}--\eqref{eq:fmunurho1} give the quantization of the moduli space of parabolic semi-stable $PSL(2,\C)$-Higgs bundles. We will check this proposal in the next section where we will compute the Hitchin character for the moduli space of parabolic semi-stable $PSL(2,\C)$-Higgs bundles on a once-punctured torus. This will give a strong check for our proposal as well as for the fusion coefficients derived above.

\subsection{$PSL(2,\mathbb{C})$ Verlinde algebra from geometry}
We can compute the metric, and fusion coefficients of the $PSL(2,\mathbb{C})$ Verlinde algebra in a different way by understanding the sign group actions on moduli stacks of flat $SL(2,\mathbb{C})$ connections on $\Sigma_{0,2}$ and $\Sigma_{0,3}$, together with their prequantum line bundles. The moduli stack of flat $SL(2,\mathbb{C})$ connections on $\Sigma_{0,3}$ is just a point, so the fusion coefficients will be 1 or 0 depending on weather the sign group action is trivial or not. The moduli stack of flat $SL(2,\mathbb{C})$ connections on $\Sigma_{0,2}$ is the groupoid $pt/\mathbb{C}^*$, so to compute the metric we have to know the sign group action on this stack, as well as the sign group action on the prequantum line.

\subsubsection*{Equivariance of the prequantum line bundle}
For a Riemann surface $\Sigma$, the $SL(2,\mathbb{C})$ prequantum line bundle assigns to a Higgs bundle $(E,\phi)$ a power of the determinant line $\mathcal{L}_{(E,\phi)} = Det^{-1}(\bar\partial_{E \otimes K^{1/2}})$. The sign group consists of line bundles $\alpha$ with $\alpha^{\otimes 2}$ identified with the trivial bundle, and acts by $E\mapsto \alpha\otimes E$. This determinant line bundle constructed from the standard representation of $SL(2,\mathbb{C})$ is not equivariant for the sign group action, but if we use the adjoint representation, it will be, because $End(E)$ is not sensitive to tensoring $E$ with line bundles. The Killing form of $SL(2,\mathbb{C})$ is 4 times the trace form of the standard representation so, when $k=0 \;\text{mod}\; 4$, the lines $Det^{-1}(\bar\partial_{End (E)\otimes K^{1/2}})^{k/4}$ will fit together to a sign group equivariant line bundle isomorphic to $\mathcal{L}^k$.

The definition of the prequantum line bundle on parabolic moduli space is more difficult, but we state here one of its features which will show up later. Choose marked points $x_1,\dots,x_n\in \Sigma$ and label them with even integers $a_1,\dots,a_n$ from $0$ to $k$. A parabolic $SL(2,\mathbb{C})$ Higgs bundle is a rank 2 Higgs bundle $(E,\phi)$ with trivialized determinant, together with a line $L_i$ in each fiber $E_{x_i}$. There is a fibration $(E,\phi,L_1,\dots,L_n)\mapsto (E,\phi)$ from parabolic moduli space to ordinary moduli space with fiber $\mathbb{C} P^1\times \dots\times \mathbb{C} P^1$. The prequantum line bundle restricted to a fiber of this fibration is isomorphic to $\mathcal{O}(a_1)\boxtimes \dots \boxtimes \mathcal{O}(a_n)$, which is the line bundle who's fiber over $(L_1,\dots,L_n)$ is $L_1^{-a_1}\otimes \dots \otimes L_n^{-a_n}$. Note that the evenness of $a_i$ is necessary for sign group equivariance. 

\subsubsection*{Geometric interpretation of the additional state}
Fixing the conjugacy class of holonomy around each boundary component of $\Sigma$ is indeed sufficient for defining a symplectic moduli space, but to define a prequantum line bundle, we need slightly more data in the non-simply connected case: we need a conjugacy class equipped with a conjugation-equivariant line bundle of prescribed curvature. The restrictions on conjugacy class and curvature come from the requirement that the line bundle trivializes the level $k$ gerbe on $G$ \cite{RISR2011gerbe}. For $PSL(2,\mathbb{C})$, there is a special conjugacy class, corresponding to $k/2$, represented by 
\[J = \begin{bmatrix}
i & 0 \\
0 & -i 
\end{bmatrix}\]
which has a disconnected point stabilizer group. To specify an equivariant line bundle on this conjugacy class, we need to choose a sign that the non-identity component of this stabilizer should act by. The two choices of this sign give rise to the states $\frac{k}{2}^{(1)}$ and $\frac{k}{2}^{(2)}$ from above. The implication for geometric quantization is that the action on the $SL(2,\mathbb{C})$ prequantum line bundle of a $\mathbb{Z}_2$ bundle which is non-trivial around a puncture labeled $\frac{k}{2}^{(i)}$ will be multiplied by $(-1)^i$. An open symmetry generating line ends at two punctures labeled $\frac{k}{2}^{(i)}$ and $\frac{k}{2}^{(j)}$, so its action on the prequantum line bundle will be multiplied by $(-1)^{i+j}$. The fact that these signs only appear in pairs is a manifestation of the dual $\mathbb{Z}_2$ symmetry of the $PSL(2,\mathbb{C})$ theory.

\subsubsection*{Fixed points of the sign group}
We can realize $\mathcal{M}_{PSL(2,\mathbb{C})}$ as a quotient of $\mathcal{M}_{SL(2,\mathbb{C})}$ by $H^1(\Sigma,\mathbb{Z}_2) = Hom(\pi_1(\Sigma),\mathbb{Z}/2)$. A fixed point of $Hom(\pi_1(\Sigma),\mathbb{Z}/2)$ is represented a homomorphism $\rho\in Hom(\pi_1(\Sigma),SL(2,\mathbb{C}))$ such that there exists $g\in SL(2,\mathbb{C})$ such that $h\rho = g\rho g^{-1}$. There is only one conjugacy class preserved under multiplication by $-1$, and it contains $J$, so if $h(\gamma)$ is non-trivial, we can conjugate $\rho$ so that $\rho(\gamma) = J$. This forces $g$ to be of the form
\[g = \begin{bmatrix}
0 & a \\
-a^{-1} & 0 
\end{bmatrix}\]
which one checks is a square root of $-I$ (and is also conjugate to $J$). In conclusion, any non-trivial sign group element will act on a bundle representing a fixed point in moduli space by a gauge transformation that is a square root of $-I$. The action of a non-trivial sign group element on the prequantum line over a fixed point in parabolic Higgs bundle moduli space will pick up a sign $(-1)^{\sum a_i/2}$ coming from $L_1^{-a_1} \otimes\dots\otimes L_n^{-a_n}$. This sign shows up when computing the $PSL(2,\mathbb{C})$ fusion coefficients $\tilde{f}^{\mu\nu a}$. 

\subsubsection*{Twice punctured sphere}
First, let $a$ be an even integer less than $k/2$. The sign group $\mathbb{Z}_2$ swaps boundary conditions $a$ and $k-a$, so the $PSL(2,\mathbb{C})$ moduli space can be identified with an $SL(2,\mathbb{C})$ moduli space.
\[\mathcal{M}_{PSL(2,\mathbb{C})}(\Sigma_{0,2},a,a) \simeq (\mathcal{M}_{SL(2,\mathbb{C})}(\Sigma_{0,2},a,a)\sqcup \mathcal{M}_{SL(2,\mathbb{C})}(\Sigma_{0,2},k-a,k-a))/\mathbb{Z}_2\]
\[ \simeq \mathcal{M}_{SL(2,\mathbb{C}}(\Sigma_{0,2},a,a)\]
so $\tilde{\eta}^{aa} = \eta^{aa}$. When $a$ is not $0$, $k$, the stack is $pt/\mathbb{C}^*$. Taking the symmetric algebra of the tangent complex, we get the Hilbert space $\mathbb{C}\oplus \mathbb{C}_t[1]$ (see \cite{Andersen:2016hoj} Theorem 20), so $\eta^{aa} = 1-t$. 

Now suppose the punctures are labeled by $\mu = \frac{k}{2}^{(i)}$ and $\nu = \frac{k}{2}^{(j)}$. The unique connection in $\mathcal{M}_{SL(2,\mathbb{C})}(\Sigma_{0,2},k/2,k/2)$ has holonomy $J$, and has automorphism group $\mathbb{C}^*$ consisting of diagonal matrices. This connection is stabilized by simultaniously multiplying by $-1$ and conjugating by an off diagonal matrix. This acts on the moduli space $pt/\mathbb{C}^*$, by inverting $\mathbb{C}^*$, so acts on the tangent complex by $-1$, thus the sign group acts on functions on the moduli space $\mathbb{C}\oplus \mathbb{C}_t[1]$ by negating $\mathbb{C}_t[1]$. The sign group action on the prequantum line comes out to be $(-1)^{i+j+k/4}$. The $PSL(2,\mathbb{C})$ Hilbert space is the $\mathbb{Z}_2$ invariant subspace. 

\[\tilde{\eta}^{\mu\nu}=
\begin{cases} 
1 & \mbox{if } (-1)^{i+j+k/4} = 1\\
-t & \mbox{else} 
\end{cases}\]

\subsubsection*{Thrice punctured sphere}
There are four cases to consider: 0,1,2 or 3 punctures labeled by $k/2$. First we compute $\tilde{f}^{abc}$ where $a,b,c\in\{0,2,..,k/2-2\}$. There are two orbits of the sign group $(\mathbb{Z}_2)^2$ action on the 8 equivalence classes of $SL(2,\mathbb{C})$ connections which map to $\mathcal{M}_{PSL(2,\mathbb{C})}(\Sigma,a,b,c)$. This gives $\tilde{f}^{abc}$ in terms of $SL(2,\mathbb{C})$ data.
\[\tilde{f}^{abc} = f^{abc} + f^{(k-a)bc}\]

Next, we compute $\tilde{f}^{ab\mu}$. The sign group acts simply transitively on the 4 relevant $SL(2,\mathbb{C})$ moduli spaces, so the $PSL(2,\mathbb{C})$ moduli space is identified with any one of these.
\[\tilde{f}^{ab\mu} = f^{ab\frac{k}{2}}\]

Next, we compute $\tilde{f}^{a\mu\nu}$ where $\mu = \frac{k}{2}^{(i)}$ and $\nu = \frac{k}{2}^{(j)}$. The sign group acts transitively on the two relevant $SL(2,\mathbb{C})$ connections, and each is stabilized by the $\mathbb{Z}_2$ bundle which is non-trivial around $x_1$ and $x_2$. The action of the sign group on the prequantum line is $(-1)^{i+j+a/2 + k/4}$.\footnote{Notice that the $k/4$ factor is necessary here to ensure the consistency in the computation for $\tilde{f}^{\mu\nu\rho}$. We hope to better understand the determinant line bundle on parabolic moduli space so as to derive it directly.} The Chern-Simons Hilbert space is $1$ dimensional if this action is trivial, and zero dimensional otherwise.
\[\tilde{f}^{\mu\nu a} = 
\begin{cases} 
1 & \mbox{if } (-1)^{i+j+a/2 + k/4} = 1\\
0 & \mbox{otherwise.} 
\end{cases}\]

Finally, we compute $\tilde{f}^{\mu\nu\rho}$ where $\mu = \frac{k}{2}^{(i)}$, $\nu = \frac{k}{2}^{(j)}$, and $\rho = \frac{k}{2}^{(l)}$. The moduli space is a single point with an action of $(\mathbb{Z}/2)^2$. The sign group action is given by the same formula as in the previous case, with $a$ set to $k/4$. For example, the action of the $\mathbb{Z}_2$ bundle which is non-trivial around $x_1$ and $x_2$ is $(-1)^{i+j}$. This gives

\[\tilde{f}^{\mu\nu\rho} = 
\begin{cases} 
1 & \mbox{if } \mu=\nu=\rho\\
0 & \mbox{otherwise.} 
\end{cases}\]

\subsection{Categorification and generalization}

The equivariant Verlinde algebra is not just an ordinary commutative Frobenius algebra, as there exist a special basis in which the fusion coefficients and the metric have non-negative integral coefficients. This is a strong hint that this algebra can be ``categorified.'' 

Indeed, physics predicts that the category of line operators $\CC_{\text{lines}}$ in the partially twisted 3d $\cN=2$ Chern--Simons--matter theory provides such a categorification. As the theory is only partially topological,  $\CC_{\text{lines}}$ is not expected to a modular tensor category when $t\neq 0$, but one expects that it still contains more data, such as braiding, compared to the commutative Frobenius algebra. 

We hope to better understand this category in future work. This will provide a systematic way of generalizing the equivariant Verlinde algebra to other groups and incorporate additional deformations.

For example, there seems to be an alternative derivation of the fusion coefficients of $\mathcal{T}(PSL(2,\C))$, that is more readily generalizable to other Lie groups. This derivation makes use of the notion of ``untwisted" and ``twisted" sector operators on which the dual $\Z_2^\zero$ acts as $+1$ and $-1$ respectively. In particular, let us consider a new basis
\bea \label{eq:basis_elements_2}
x^0 , x^2, \cdots, x^{\frac{k}{2}-2}, x^+, x^-
\eea
where $x^0 , x^2, \cdots, x^{\frac{k}{2}-2}$, and $ x^+ = x^{\frac{k}{2}}$ are the ``untwisted" sector operators and 
\bea 
x^- = x^{{\frac{k}{2}}^\one} - x^{{\frac{k}{2}}^\two}
\eea
is the ``twisted" sector operator.
This operator may be thought of as living at the end of an open $\Z_2^\zero$ symmetry generating line in the parent TQFT $\CT(SL(2,\C)_k)/\Z_2^\one$. It carries charge $-1$ under the dual $\Z_2^\zero$ symmetry.

The fusion coefficients involving the untwisted sector operators are fixed entirely through some straightforward algebra. In particular, $\widetilde{f}^{++}{}_a = 2 f^{\frac{k}{2}\frac{k}{2}}{}_a$. 
 Invariance under the dual $\Z_2^\zero$ symmetry implies that fusion coefficients involving $x^-$ are non-zero only if $x^-$ appears an even number of times. For example, $\widetilde{f}^{-a}{}_b =0$. Then, we have the relation
\bea \label{eq:relation}
\widetilde{f}^{- -}{}_a = (-1)^{ \frac{a}{2}+\frac{k}{4}} \widetilde{f}^{++}{}_a, 
\eea
which can be used to completely fix all the rest of the fusion coefficients without using positivity and integrality. At the same time, it impose a very strong condition on the other fusion coefficients as $a$ is arbitrary.   
The phase $(-1)^{ \frac{a}{2}+\frac{k}{4}}$ in the $t\to 0$ limit can be interpreted as a combination of R- and F-matrices, and has a generalization to other Lie groups. Understanding whether and why such structure preserve when $t$ is non-zero would shed light on the structure of the  category $\CC_{\text{lines}}$, which we hope to investigate in the future.

\section{Quantization of $\cM(PSL(2,\mathbb{C}))$ for once-punctured torus} \label{sec:parabolic_torus}
In this section, we will compute the equivariant trace and the Hitchin character for the moduli space of parabolic Higgs bundles on a once-punctured torus. First, we will compute the equivariant trace in Section \ref{sec:torusfromtqft} using methods in 2d TQFT. Then in Section \ref{sec:once-punctured}, we will propose a method for computing this result directly from geometry using the Atiyah-Singer-Lefschetz index theorem. Finally, in Section \ref{sec:eqvftorus}, we will compute the Hitchin character of $\cM_{w_2}(PSL(2,\C))$, and show that it is consistent with $PSL(2,\C)$ equivariant Verlinde algebra studied in Section \ref{sec:PSL2Calgebra}.

\subsection{Computation from 2d TQFT} \label{sec:torusfromtqft}

Let $\cM^a_d$ be the moduli space of parabolic semi-stable $SL(2,\C)$-Higgs bundles of degree $d$ on a once-punctured torus where $a$ denotes a highest weight of $\mathfrak{su}(2)$ that specifies the parabolic structure at the puncture. We will set $k \equiv 0 \pmod 4$ and $a \equiv 0 \pmod 2$. $\cM^a_{d}$ admits an action of $H_1 (\Sigma_{1,1}, \Z_2)$. The equivariant trace of an element  $h \in H_1(\Sigma_{1,1}, \Z_2)$ can be computed using the methods of Section 2. For any non-zero element $h$, $\Trt (h|H^0 (\cM_{d}, \cL^k))$ gives the partition function of $\CT(SL(2,\C)_k)$ in the presence of a zero-form symmetry background. This partition function is the same for all non-zero $h$. 

\begin{figure}
\centering
        \includegraphics[totalheight=4cm]{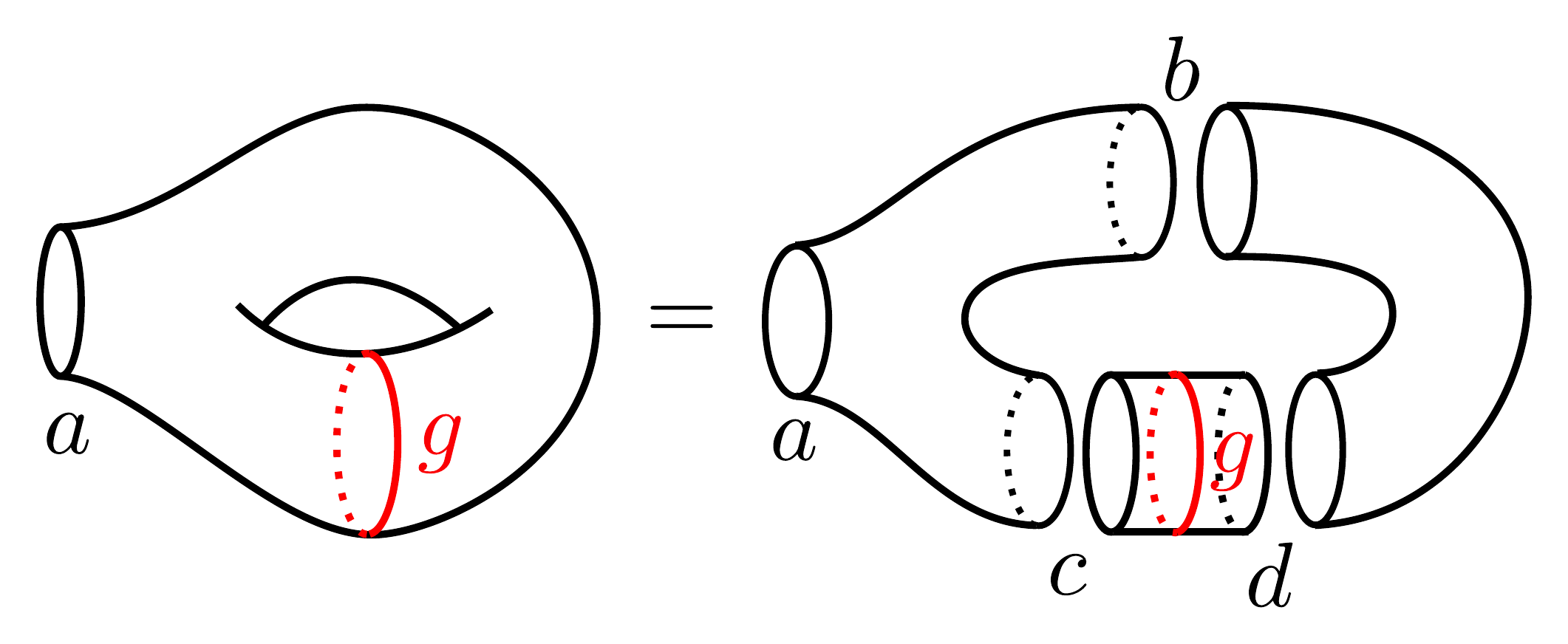}
    \caption{Once-punctured torus}
    \label{fig:torus}
\end{figure}

Let us consider the configuration in \autoref{fig:torus}. Here, the twisted cylinder has components $\overline{\eta}^a{}_b = (-1)^{a} \delta^a{}_b$, and we obtain
\bea \label{eq:trS1a}
\Trt (h|H^0 (\cM_{d=0}, \cL^k)) = \sum\limits_{b,c,d} f^a{}_{bc} \overline{\eta}^c{}_d \eta^{bd} = \sum\limits_{b} (-1)^b f^{abb} \eta_{bb}.
\eea
Using the fusion rules of $SL(2,\C)$ (4.1)--(4.2), we get
\bea 
f^{abb} =  
\begin{cases} 
1 &\mbox{if } \frac{a}{2} \leq b \leq k-\frac{a}{2} \\
t^{c/2} &\mbox{if } b<\frac{a}{2} \ \text{or} \ b> k- \frac{a}{2} 
\end{cases}
\eea
where $c=\text{max}(\frac{a}{2}-b, \frac{a}{2}+b-k)$. Therefore,
\bea \nonumber
\Trt (h|H^0 (\cM_{d=0}, \cL^k)) &=& \sum\limits_{b=0}^k (-1)^b f^{abb} \eta_{bb} \\ \nonumber
&=& \frac{2t^{a/2}}{1-t^2} +  \frac{2 t^{\frac{a}{2}}}{1-t} \sum\limits_{b=1}^{\frac{a}{2}-1} (-t)^{-b}+ \sum\limits_{b=\frac{a}{2}}^{k-\frac{a}{2}} \frac{(-1)^b}{1-t} \\ \nonumber
&=& \frac{2t^{a/2}}{1-t^2} + \frac{2t^{a/2}}{1-t^2} \sum\limits_{b=1}^{\frac{a}{2}-1} (-1)^b (t^{-b} + t^{-b+1}) + \frac{(-1)^{a/2}}{1-t} \\ \nonumber
&=& \frac{2t (-1)^{a/2-1}}{1-t^2} +\frac{(-1)^{a/2}}{1-t} \\ \label{eq:tra>0}
&=& \frac{(-1)^{a/2}}{1+t}.
\eea

For $d=1$, we introduce a $\Z_2^\one$ background labelled by the generator of $\Z_2^\one$, i.e. $(w^k - tw^{k-2})$. This operator fuses with $w^a$ to give $w^{k-a}$. The equivariant trace is obtained by replacing $a$ with $k-a$. Since $k\equiv 0 \pmod 4$, this does not change the result, and we get
\bea \label{eq:d=1_case_torus}
\Trt (h|H^0 (\cM_{d=1}, \cL^k)) = \frac{(-1)^{a/2}}{1+t}.
\eea

\subsection{Equivariant trace: a geometric computation} \label{sec:once-punctured}
For arbitrary simple complex Lie groups, the equivariant trace and the Hitchin characters of $\cM_{w_2} (G)$ may be computed using a trace formula due to Atiyah and Singer \cite{10.2307/1970717}. Let $W$ be a holomorphic vector bundle on $Y=X/\CG$ where $X$ is a compact complex manifold and $\CG$ is a finite group.
Let $V$ be the pullback of $W$ on $X$, then the Euler characteristic $\chi (Y,W) := \sum\limits_p (-1)^p \text{dim} \ H^p (Y|\CO(W))$ satisfies
\bea \label{eq:Euler}
\chi (Y,W) = \frac{1}{|\CG|} \sum\limits_{h\in \CG} \mu (h|X,V) 
\eea 
where 
\bea \label{eq:define_mu_h}
\mu (h|X, V) := \sum\limits_{p} (-1)^p \Tr (h|H^p (X|\CO(V)))
\eea
 can be computed as an integral over the fixed point locus $P$ of $h$,
\bea \label{eq:Lefschetz_trace}
\mu (h|X, V)= \int_P \frac{\text{Td}(T_P) \widetilde{\text{ch}}(V_P, h)}{\lambda(N_{P}, h)}.
\eea

Here $\text{Td}(T_P)$ is the Todd class of $P$, and $\widetilde{\text{ch}}(V_P,h)$ and $\lambda (N_{P}, h)$ are defined as follows. Let $V_P$ be the restriction of $V$ to $P$; and let $\mu_i$ be the eigenvalues of $h$ as an endomorphism of $V_P$. Then, with the eigen-decomposition $V_P = \oplus_i V_{\mu_i}$ we define 
\bea 
\widetilde{\text{ch}}(V_P,h) := \sum\limits_i \mu_i \text{ch}(V_{\mu_i}).
\eea 
Similarly, the normal bundle $N_{P}$ of $P$ in $X$ also has an eigen-decomposition, $N_{P} = \oplus_i N_{\lambda_i}$ with eigenvalues $\lambda_i$ of $h$, and we define
\bea \label{eq:lambda}
\lambda (N_{P}, h) = \prod\limits_i \sum\limits_{p\geq 0} (-\lambda_i)^p \text{ch} (\Lambda^p N_{\lambda_i}^\star).
 \eea
 For $h=e$, $\mu(h|X,V)$ reduces to the usual Riemann--Roch formula that counts the dimension of the space of sections of $V$. 

We wish to use this general result to compute the equivariant trace $\Trt (h|H^0 (\cM_{d}, \cL^k))$ and the Hitchin character $\text{dim}_t H^0 (\cM_{w_2}(PSL(2,\C)), \cL^k)$. Two issues appear at once.
\begin{enumerate}
\item $\mu(h|X,V)$ computes an Euler characteristic while the equivariant trace $\Trt (h|H^0 (\cM_{d}, \cL^k))$ is computed on the zeroth cohomology, and
\item The theorem above required $X$ and $Y$ to be compact manifolds, while $\cM_{w_2}(PSL(2,\C))$ is non-compact.
\end{enumerate}

The first issue is resolved by using the fact that the higher cohomology $H^i (\cM_{d}, \cL^k)$ for $i>0$ vanishes. This result was proved in \cite{Andersen:2016hoj} and \cite{halpernleistner2016equivariant}. The second issue can be cured by considering a $\C^\star$-equivariant version of the theorem. Alternatively, one may use the identification \cite{Andersen:2016hoj,halpernleistner2016equivariant} 
\bea
H^i_n (\cM,\cL^k) = H^i (\mathfrak{N},  S^n T\mathfrak{N} \otimes \cL^k)
\eea
to transfer the computation to the moduli stack $\mathfrak{N}$ of $G$-\textit{vector} bundles. This gives
\bea \label{eq:char_in_terms_of_N}
\Trt (h|H^0 (\cM_{d=w_2}, \cL^k)) = \sum\limits_{n=0}^\infty \mu(h| \mathfrak{N},  S^n T\mathfrak{N} \otimes \cL^k) t^n.
\eea
Each summand on the right side can be computed using equation \eqref{eq:Lefschetz_trace}.

In what follows, we will propose that in the case of once-punctured torus, the difference between the moduli stack and the moduli space $\CN$ of semi-stable $G$ vector bundles will not matter. That is,
\bea 
 \mu(h| \mathfrak{N},  S^n T\mathfrak{N} \otimes \cL^k)=  \mu(h| \mathcal{N},  S^n T\mathcal{N} \otimes \cL^k)
\eea 
when $G=SL(2,\C)$ and the underlying surface is the once-punctured torus. Our proposal is based on the observation that 
\bea \label{eq:moduli_space_trace}
 \sum\limits_{n=0}^\infty \mu(h| \mathcal{N},  S^n T\mathcal{N} \otimes \cL^k) t^n
 \eea
 is equal to the equivariant trace computed in the previous section. This is interesting because while the higher cohomology $H^{i>0} (\mathfrak{N},  S^n T\mathfrak{N} \otimes \cL^k)$ vanishes for the moduli stack  \cite{Andersen:2016hoj,halpernleistner2016equivariant}, it does not vanish for the moduli space. Furthermore, the zeroth cohomology on the stack and the moduli space coincide. Therefore, our proposal implies in particular that $\Tr (h|H^{i>0} (\cN,S^n T\cN \otimes \cL^k)) = 0$.

In the rest of this section, we will compute \eqref{eq:moduli_space_trace} for once-punctured torus and $G=SL(2,\C)$.
 Through the Narasimhan--Seshadri theorem, $\cN$ is also the moduli space of $SU(2)$ flat connections with the holonomy around the puncture conjugate to $e^{i \pi  \frac{a}{k} \sigma_3}$.  
This moduli space is a closed surface inside $\mathbb{R}^3$, and is described by the equation
\cite{Gukov:2010sw}
\bea \label{eq:moduli_space}
x^2 + y^2 + z^2 - xyz = 4 \cos^2 \left( \frac{\pi a }{2k} \right) 
\eea
where $x,y,z \in [-2,2]$.  $H_1 (\Sigma_{1,1}, \Z_2) = \Z_2 \times \Z_2$ acts on this surface as $(x,y,z) \to (\pm x, \pm y, \pm z)$ with an even number of minus signs. A non-zero element of this group has two fixed points, which we will denote by $p_1$ and $p_2$. The integral in equation \eqref{eq:Lefschetz_trace} reduces to a sum,
\bea \label{eq:sum_over_points}
\mu (h | \cN_a, S^n T\cN_a \otimes \cL^k )= 
\sum\limits_{i=1}^2 \frac{\text{Td}(T_{p_i}) \ \widetilde{\text{ch}}(V_{p_i}, h)}{\lambda (N_{p_i}, h)}.
\eea
As $h$ acts on the normal bundle $N_{p_i}$ with eigenvalue $-1$, $\lambda(N_{p_i}, h) = \text{ch} (\Lambda^0 N^\star_{p_i}) + \text{ch} (\Lambda^1 N^\star_{p_i}) $. After restricting to a point, we obtain $\lambda(N_{p_i}, h) =2$. 

To compute $\widetilde{\text{ch}}(V_{p_i},h)$ for $V=S^n (T\mathbb{CP}^1) \otimes \cL^k$, we note that as $N_{p_i} \equiv T\mathbb{CP}^1_{p_i}$, $h$ acts on the tangent space $T\mathbb{CP}^1_{p_i}$ by multiplication by $-1$. It therefore acts on $\cL^k \equiv (T\mathbb{CP}^1)^{\frac{k-a}{2}}$ by $(-1)^{a/2}$ (as $k\equiv 0 \pmod 4$). Overall, $
\widetilde{\text{ch}}(V_{p_i},g) = (-1)^{n+\frac{a}{2}} \text{ch} (V_{p_i}) $, and
\bea 
\mu(h| \mathcal{N},  S^n T\mathcal{N} \otimes \cL^k) = 
\sum\limits_{i=1}^2 \frac{(-1)^{n+\frac{a}{2}}}{2} = (-1)^{n+\frac{a}{2}}.
\eea
This gives
\bea \label{eq:geometric_answer}
\sum\limits_{n=0}^\infty \mu(h| \mathcal{N},  S^n T\mathcal{N} \otimes \cL^k) t^n  =  \frac{(-1)^{a/2}}{1+t}
\eea
which is precisely the result \eqref{eq:tra>0} obtained in the previous subsection.

Finally, we remark that the case $d=1$ can be treated by introducing another puncture on the torus with holonomy $-\mathbb{1}$. This effectively replaces $a$ with $k-a$ in equation \eqref{eq:moduli_space}. Making this transformation in \eqref{eq:geometric_answer} leaves the right hand side invariant, and we get
\bea \label{eq:d=1_case}
\sum\limits_{n=0}^\infty \mu(h| \mathcal{N},  S^n T\mathcal{N} \otimes \cL^k) t^n   =  \frac{(-1)^{a/2}}{1+t}.
\eea
which is equal to the result \eqref{eq:d=1_case_torus} computed in the previous subsection.

\subsection{Hitchin characters for $\mathcal{M}_{w_2}(PSL(2,\mathbb{C}))$} \label{sec:eqvftorus}

In this section, we will use the formulae for equivariant trace derived in the previous two sections to compute the Hitchin characters for $\cM_{w_2} (PSL(2,\C))$. Let us first consider the case $a \in \{ 0, 2, \cdots, \frac{k}{2}-2\}$. The case of $a=\frac{k}{2}$ will be discussed shortly. For $w_2 = 0$, 
\bea \label{eq:ZaPSL2Cw0} 
\text{dim}_t H^0 (\cM^a_{w_2=0}(PSL(2,\C)), \cL^k) = \frac{1}{4} \sum\limits_{h \in H_1(\Sigma_{1,1},\Z_2)}  \Trt (h|H^0 (\cM^a_{d=0}, \cL^k))
\eea
For $h=0$, $\Trt (h|H^0 (\cM^a_{d=0}, \cL^k))$ is the Hitchin character for $\cM_{d=0}$, which can be computed using the fusion rules in Section \ref{sec:PSL2Calgebra},
\bea 
\text{dim}_t H^0 (\cM^a_{d=0}, \cL^k) = \frac{k-a+1}{1-t} + \frac{2t}{(1-t)^2} + \frac{4t^{a/2}}{(1-t)(1-t^2)}
\eea
This gives
\bea \nonumber \label{eq:explicit_ZaPSL2Cw0}
\text{dim}_t H^0 (\cM_{w_2=0}(PSL(2,\C), \cL^k) &=& \frac{1}{4}\left( \frac{k-a+1}{1-t} + \frac{2t}{(1-t)^2} + \frac{4t^{a/2}}{(1-t^{-1})(1-t^2)} +\frac{3 (-1)^{a/2}}{1+t}\right). \\
\eea
The Hitchin character for $w_2=1$ component can be obtained by replacing $a$ with $k-a$, 
\bea \label{eq:explicit_ZaPSL2Cw1} \nonumber
\text{dim}_t H^0 (\cM_{w_2=1}(PSL(2,\C), \cL^k) &=& \frac{1}{4}\left( \frac{a+1}{1-t} + \frac{2t}{(1-t)^2} + \frac{4t^{(k-a)/2}}{(1-t^{-1})(1-t^2)} +\frac{3 (-1)^{a/2}}{1+t}\right). \\
\eea

The union $\cup_{w_2} \cM_{w_2} (PSL(2,\C))$ is the classical phase space of $PSL(2,\C)$ Chern--Simons theory. After $\beta$-deformation, the partition function of this theory on $S^1 \times \Sigma_{1,1}$ is the sum of \eqref{eq:explicit_ZaPSL2Cw0} and \eqref{eq:explicit_ZaPSL2Cw1}. Let us denote the sum by $\CZ_{\CT(PSL(2,\C))}$, then it is easy to check that
\bea 
\CZ_{\CT(PSL(2,\C))} = \sum\limits_{\substack{b=0 \\ b \ even}}^{\frac{k}{2}-2} \widetilde{f}^{ab}{}_b + \sum\limits_{\mu \in \{\frac{k}{2}^\one, \frac{k}{2}^{(2)}\}} \widetilde{f}^{a\mu}{}_\mu 
\eea
where $\widetilde{f}^{ab}{}_b$ and $\widetilde{f}^{a\mu}{}_\mu $ are the fusion coefficients of $PSL(2,\C)$ Verlinde algebra in Section \ref{sec:PSL2Calgebra}.

Now let $a=\frac{k}{2}$. There are two operators in the TQFT labelled by this weight, and the dual $\Z_2^\zero$ symmetry exchanges them. The geometric interpretation of these two boundary conditions is not clear, so we will suffice ourselves with the physical interpretation of the partition functions. Let $\CZ^{(1)}$ and $\CZ^\two$ be the partition functions of $\CT(SL(2,\C)_k)/\Z_2^0$, then the dual $\Z_2^\zero$ symmetry implies that
\bea 
\CZ^{(1)} = \CZ^{(2)}. 
\eea
If the puncture on the torus is labelled by $w^{\frac{k}{2}}=w^{\frac{k}{2}^\one}+w^{\frac{k}{2}^\two}$, we can use equation
\bea 
\CZ^{(1)} + \CZ^{(2)} = \frac{1}{2} \sum\limits_{h \in H_1(\Sigma_{1,1},\Z_2)} \CZ^{\frac{k}{2}}_h
\eea
where
\bea
\CZ^{\frac{k}{2}}_h =
\begin{dcases} 
 \frac{k+2}{2(1-t)} + \frac{2t}{(1-t)^2} + \frac{4t^{k/4}}{(1-t^{-1})(1-t^2)} 
 &\mbox{if } h= 0 \\ 
\frac{(-1)^{k/4}}{1+t} &\mbox{if } h \neq 0 
\end{dcases}
\eea
 to obtain
\bea \nonumber
\CZ^{(1)} = \CZ^{(2)} = \frac{1}{4}\left( \frac{k+2}{2(1-t)} + \frac{2t}{(1-t)^2} + \frac{4t^{k/4}}{(1-t^{-1})(1-t^2)} +\frac{3 (-1)^{k/4}}{1+t} \right).
\eea

Introducing a one-form symmetry background replaces $a$ with $k-a$ but for $a=\frac{k}{2}$, this has no effect. The partition function of $\CT(PSL(2,\C)_k)$ labelled by the two boundary conditions $\frac{k}{2}^{(i)}$ is
\bea \label{eq:Zk/2PSL2C}
\CZ^{(i)}_{\CT(PSL(2,\C))} = \frac{1}{4}\left( \frac{k+2}{2(1-t)} + \frac{2t}{(1-t)^2} + \frac{4t^{k/4}}{(1-t^{-1})(1-t^2)} +\frac{3 (-1)^{k/4}}{1+t} \right).
\eea
This partition function is also reproduced by the fusion coefficients of $PSL(2,\C)$ algebra studied in Section \ref{sec:PSL2Calgebra},
\bea \nonumber
\CZ^{(i)}_{\CT(PSL(2,\C))} &=& \sum\limits_{a} \widetilde{f}^{\frac{k}{2}^{(i)}a}{}_a + \sum\limits_\mu \widetilde{f}^{\frac{k}{2}^{(i)}\mu}{}_\mu .
\eea
giving another check on the formulae obtained there.

\subsection*{Acknowledgements}
We would like to thank Dan Freed, Po-Shen Hsin, Andrew Neitzke, and Sebastian Schulz for useful discussions. We would also like to thank Simons Center for Geometry and Physics for generous hospitality during Graduate Summer School on the Mathematics and Physics of Hitchin Systems and the Simons Summer Workshop 2021. 
The work of SG is supported by the U.S. Department of Energy,
Office of Science, Office of High Energy Physics, under Award No. DE-SC0011632, and by
the National Science Foundation under Grant No. NSF DMS 1664227. 
The work of AS was supported by
NSF grants DMS-2005312 and DMS-1711692. 
The work of DP was supported by the Center for Mathematical Sciences and Application. This paper is partly a result of the ERC-SyG project, Recursive and Exact New Quantum Theory (ReNewQuantum) which received funding from the European Research Council (ERC) under the European Union's Horizon 2020 research and innovation programme under grant agreement No 810573.

\appendix
\section{Partition function of $\mathcal{T}/\mathbb{Z}_n^{(0)}$} \label{sec:part_func}
The partition function of $\CT/\Z_n^\zero$ is defined as
\bea \label{eq:Z_orig_app}
\CZ_{\CT/\Z_n^\zero} = \frac{1}{|H^1(\Sigma, \Z_n)|^{\frac{1}{2}}} \sum\limits_{h \in H_1 (\Sigma, \Z_n)} \CZ_{h}.
\eea
We will show in Appendix \ref{sec:comp_Zh} that for $h \neq 0$,
\bea \label{eq:Z_h_app}
\CZ_{h} = \sum\limits_{i \in \CB^{\Z_m}}  N_i^{2g-2}
\eea
where $\Z_m$ is the smallest subgroup of $\Z_n$ such that $h \in H_1(\Sigma, \Z_m)$, and $\CB^{\Z_m}
\subset \CB$ is the set of $\Z_m$-fixed points. Using this result, we will show in Appendix \ref{sec:ZT/Z_n0} that
\bea \label{eq:Z_main_app}
\CZ_{\CT/\Z_n^\zero} &=& \sum\limits_{i \in \CB/\Z_n}|\Stab(i)| \left( \frac{|\Stab(i)|}{|\Z_n|^\half}  N_i \right)^{2g-2}
\eea
where $\CB/\Z_n$ is the set of $\Z_n$-orbits of $\CB$, and for any orbit $i \in \CB/\Z_n$, the cardinality of the stabilizer of any of its representatives in $\CB$ is denoted by $\{1,2,\cdots, |\Stab (i)|\}$. Finally, in Appendix \ref{sec:product} we will discuss gauging a product of cyclic groups.

\subsection{Computation of $\mathcal{Z}_h$} \label{sec:comp_Zh}
The main idea behind the derivation of equation \eqref{eq:Z_h_app} is that for any $\Z_n$ bundle over a surface, the surface can be chopped into pairs of pants and cylinders such that the restriction of the $\Z_n$ bundle to each pair of pants is trivial, but the restriction to the cylinders may not be. In TQFT, this means that the only new ingredients required for the computation of $\CZ_h$ are the ``twisted cylinders". These enact the $\Z_n$ action on $\CB$ giving $\eta^{i}{}_j = \delta^i{}_{z(i)}$ where $z\in \Z_n$. Now gluing the surface back together using the twisted cylinders enforces the sum over the diagonal basis $\CB$ to collapse to the subset that is invariant under the action of the twisted cylinders. We will now make this idea more concrete. 

\begin{figure}
\centering
        \includegraphics[totalheight=3cm]{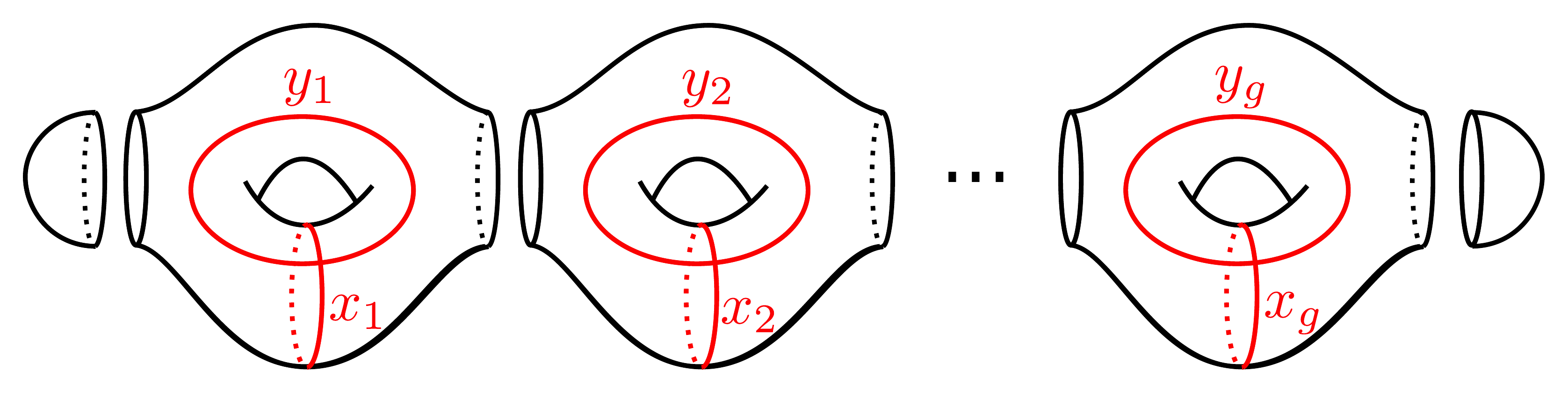}
    \caption{Decomposition of $\Sigma$ into caps and handles}
    \label{fig:genusg}
\end{figure}

\begin{figure}
\centering
        \includegraphics[totalheight=4cm]{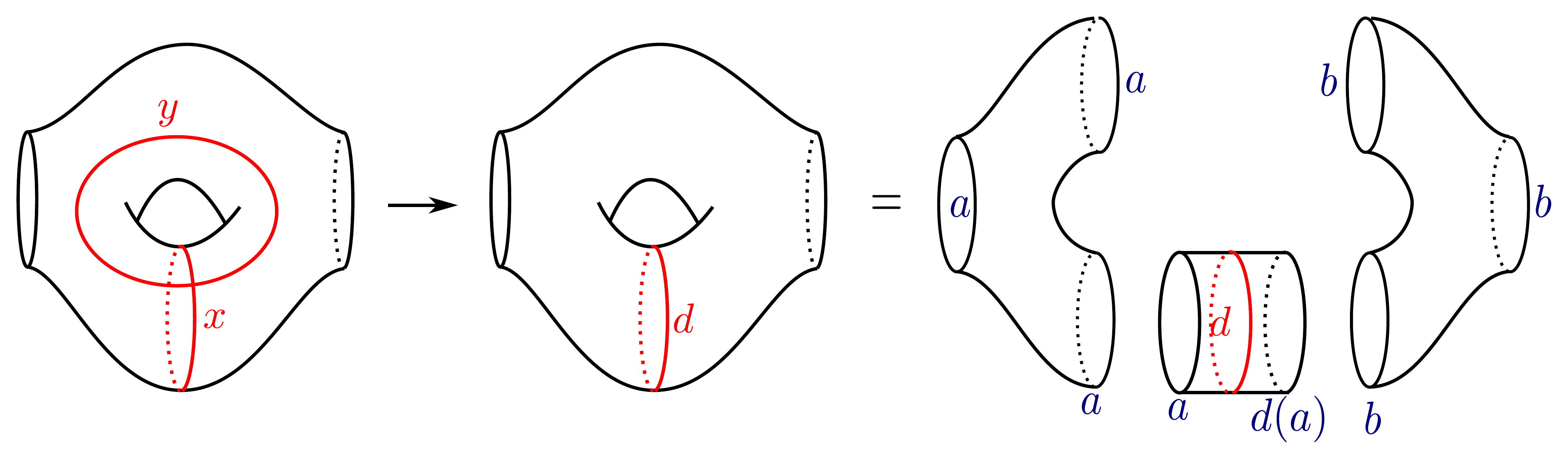}
    \caption{A handle operator decorated with $(x_i,y_i) \in \Z_n \times \Z_n$. Under the action of modular group, it can be mapped to a handle operator with decoration $(\text{gcd}(x_i,y_i),0)$. Finally, consistent gluing requires that $a=d_i(a)=b$, which implies that the handle operator is non-zero only if this condition is satisfied.}
    \label{fig:handle}
\end{figure}

Consider a standard basis $\{A_i, B_i; i = 1, \cdots, g\}$ of cycles in $H_1 (\Sigma, \Z_n)$ and write $h = \sum\limits_{i=1}^g x_i A_i + y_i B_i$ for $x_i, y_i \in \Z_n$. It is convenient to chop the surface into handles and caps as in \autoref{fig:genusg}, and apply the action of the mapping class group of each handle to bring $h$ to a form $\sum\limits_{i=1}^g d_i A_i$ where $d_i =\text{gcd}(x_i, y_i)$.\footnote{The relevant element of $SL(2,\Z)$, i.e.  
$\begin{pmatrix}
a & b \\ c & e \end{pmatrix}$ is such that that $c = - \frac{y}{d}$, $e = \frac{x}{d}$, and $(a,b)$ is a Bezout's pair satisfying $a x + b y= d$.} (See Figure \ref{fig:handle}). Figure \ref{fig:handle} also shows that the handle operator labelled by $(x_i, y_i)$ has a non-zero matrix entry only if the two boundaries are labelled by the same $d_i$-invariant state. Gluing the handle operators and caps back together we find 
\bea 
\CZ_h = \sum\limits_{i \in \CB^{\Z_m}} N_i^{2g-2}
\eea
where $\Z_m$ is the smallest subgroup containing all $\text{gcd}(x_i,y_i)$. As the action of mapping class group preserves each subgroup of $H_1(\Sigma, \Z_n)$, $\Z_m$ is also the smallest subgroup such that $H_1(\Sigma, \Z_m)$ contains $h$. This completes the derivation of equation \eqref{eq:Z_h_app}.

\subsection{Computation of $\mathcal{Z}_{\CT/\Z_n^\zero}$} \label{sec:ZT/Z_n0}
Now we will derive the result \eqref{eq:Z_main_app}. The derivation is straightforward when $n=p$ is prime, so we discuss this case first. We recall that $\Z_p$ has no non-trivial subgroups. Therefore, $\CZ_h$ is the same for all non-zero $h$, and we get
\bea \nonumber \label{eq:0FormGauged_n_prime}
\CZ_{\CT/\Z_p^\zero} &=& \frac{1}{p^g} \left( \sum\limits_{i \in \CB} N_i^{2g-2} + (p^{2g}-1) \sum\limits_{i\in \CB^{\Z_p} } N_i^{2g-2} \right) \\ \nonumber
&=& \frac{1}{p^g} \left( \sum\limits_{\substack{i \in \CB \backslash \CB^{\Z_p}}} N_i^{2g-2} + p^{2g} \sum\limits_{i\in \CB^{\Z_p} } N_i^{2g-2} \right)\\ \nonumber
&=& \sum\limits_{\substack{i\in \CB/\Z_p\\ |i|=p}} \left(\frac{N_i}{|\Z_p|^\half}\right)^{2g-2} +
p^{2g-1} \sum\limits_{\substack{i\in \CB/\Z_p\\ |i|=1}}  N_i^{2g-2}  \\
 &=& \sum\limits_{i \in \CB/\Z_p}|\Stab(i)| \left( \frac{|\Stab(i)|}{|\Z_p|^\half}  N_i \right)^{2g-2}
\eea
This is precisely the result \eqref{eq:Z_main_app}.

When $n$ is composite, $\Z_n$ has a non-trivial subgroups which are labelled by divisors of $n$, and if $H_1(\Sigma, \Z_m)$ is the smallest subgroup containing $h$ then $\CZ_h$ gets contribution only from $\Z_m$-invariant elements of $\CB$. 

In order to evaluate \eqref{eq:Z_orig_app} we need to know for each $m | n$, the number of elements of $H_1(\Sigma, \Z_m)$ that are not contained in any smaller subgroup. This number is computed as follows. Let $n =p_1^{a_1} \cdots p_j^{a_j}$ be the prime factorization of $n$, then a divisor $m$ of $n$ has the prime factorization $m = p_1^{b_1} \cdots p_j^{b_j}$ where $0 \leq b_i \leq a_i$. In particular, we may represent a divisor of $n$ by a tuple $(b_1, \cdots, b_j)$ such that $0 \leq b_i \leq a_i$, and denote the quantities depending on a divisor accordingly. For example, let  $\Z_{(b_1, \cdots, b_j)}$ be another notation for $\Z_{p_1^{b_1} \cdots p_j^{b_j}}$, $\CB^{(b_1, \cdots, b_j)}$ denote the set of elements fixed under the action of $\Z_{(b_1, \cdots, b_j)}$, and so on.

The number of elements of $H_1 \left(\Sigma, \Z_{p_1^{b_1} \cdots p_j^{b_j}}\right)$ that are not contained in any smaller subgroup is 
\bea \nonumber \label{eq:valueofC}
\CC_{(b_1, \cdots, b_j)} &=& 
\left(\prod\limits_{ q =1}^j p_q^{b_q}\right)^{2g} - \sum\limits_{r=1}^j \left( p_r^{b_r-1} \prod\limits_{ q \neq r} p_q^{b_q} \right)^{2g} + \sum\limits_{1 \leq r < s < j} 
 \left( p_r^{b_r-1}  p_s^{b_s-1} \prod\limits_{ \substack{q \neq r \\ q \neq s}} p_q^{b_q} \right)^{2g} 
 \\ 
&&- \cdots + (-1)^{j-1} \left(\prod\limits_{ q =1}^j p_q^{b_q-1} \right)^{2g}.
\eea
This result follows from a use of the inclusion-exclusion principle: the first term is the number of elements in $H_1 (\Sigma, \Z_{(b_1, \cdots, b_j)})$; the second term is the number of elements in a subgroup $H_1 (\Sigma, \Z_{(b_1, \cdots,b_r - 1, \cdots, b_j)})$; the third term includes again the number of elements in the pairwise intersections of $H_1 (\Sigma, \Z_{(b_1, \cdots,b_r - 1, \cdots, b_j)})$ and $H_1 (\Sigma, \Z_{(b_1, \cdots,b_q - 1, \cdots, b_j)})$, and so on. 

Using \eqref{eq:valueofC}, we obtain the following form for the sum in \eqref{eq:Z_orig_app},
\bea \label{eq:partition_function_2}
 \sum\limits_{h \in H_1 (\Sigma, \Z_n)} \CZ_h = \sum\limits_{\substack{(b_1, \cdots, b_j) \\ 0 \leq b_j \leq a_j}} \CC_{(b_1, \cdots, b_j)} \sum\limits_{i \in \CB^{(b_1, \cdots, b_j)}} N_i^{2g-2}.
\eea
The coefficient of $\left(\prod\limits_{ q =1}^j p_q^{b_q}\right)^{2g}$ in this sum gets a contribution from a tuple $(c_1, \cdots, c_j)$ if for all $i\in \{1,\cdots, j\}$, $c_i = b_i$ or $c_i = b_i + 1$. In particular, the coefficient of $\left(\prod\limits_{ q =1}^j p_q^{b_q}\right)^{2g}$ is
\bea \nonumber \label{eq:sum_over_Fs}
\sum\limits_{i \in \CB^{(b_1, \cdots, b_j)}} N_i^{2g-2} &&-
\sum\limits_{r=1}^j \sum\limits_{i \in \CB^{(b_1, \cdots, b_r+1 , \cdots, b_j)}} N_i^{2g-2} + 
\sum\limits_{1 \leq r < s \leq j} \sum\limits_{i \in \CB^{(b_1, \cdots, b_r+1 , \cdots, b_{s+1}, \cdots, b_j)}} N_i^{2g-2} \\
&&- \cdots + (-1)^{j-1}  \sum\limits_{i \in \CB^{(b_1+1, \cdots, b_r+1 , \cdots, b_{s+1}, \cdots, b_j+1)}} N_i^{2g-2}.
\eea

This expression is equal to the sum of $N_i^{2g-2}$ over the elements of $\CB$ that lie in a $\Z_n$ orbit of length $\prod\limits_{q=1}^j p_q^{a_q-b_q}$.\footnote{This can be seen as follows. 
The first sum in \eqref{eq:sum_over_Fs} is the sum over elements fixed under $\Z_{p_1^{b_1} \cdots p_j^{b_j}}$. Such an element lies in a $\Z_n$ orbit of length $p_1^{x_1} \cdots p_j^{x_j}$ where $x_j \leq a_j - b_j$. Using inclusion-exclusion principle, we can remove the set of elements with $x_j < a_j - b_j$. This is precisely what the rest of the terms in \eqref{eq:sum_over_Fs} accomplish.} 
Let $\Orb_{(b_1, \cdots, b_j)}(\CB)$ be the set of orbits of $\CB$ of this length, then the expression \eqref{eq:sum_over_Fs} is equal to 
\bea 
 \prod\limits_{q=1}^j p_q^{a_q-b_q} \sum\limits_{i \in \Orb_{(b_1, \cdots, b_j)}}  N_i^{2g-2} 
\eea
Plugging this in equation \eqref{eq:Z_orig_app}, we get
\bea \nonumber
\CZ_{\CT/\Z_n^\zero} &=& \frac{1}{n^g} \left( \sum\limits_{\substack{(b_1, \cdots, b_j) \\ 0 \leq b_j \leq a_j}}  \left(\prod\limits_{ q =1}^j p_q^{b_q}\right)^{2g} 
\prod\limits_{q=1}^j p_q^{a_q-b_q} 
\sum\limits_{i \in \Orb_{(b_1, \cdots, b_j)}}  N_i^{2g-2} 
 \right) \\ \nonumber
&=& \sum\limits_{\substack{(b_1, \cdots, b_j) \\ 0 \leq b_j \leq a_j}} \left(\prod\limits_{ q =1}^j p_q^{b_q}\right)^{2g-1} \sum\limits_{i \in \Orb_{(b_1, \cdots, b_j)}(B)} \left( \frac{N_i}{|\Z_n|^\half}\right)^{2g-2} \\ \nonumber
&=&  \sum\limits_{\substack{(b_1, \cdots, b_j) \\ 0 \leq b_j \leq a_j}}
 \sum\limits_{i \in \Orb_{(b_1, \cdots, b_j)} (B)}
 \left(\prod\limits_{ q =1}^j p_q^{b_q}\right)^{2g-1} \left( \frac{N_i}{|\Z_n|^\half}\right)^{2g-2} \\
 &=& \sum\limits_{i \in \CB/\Z_n} |\Stab(i)| \left( \frac{|\Stab(i)|}{|\Z_n|^\half}  N_i \right)^{2g-2}
\eea
This completes the derivation of equation \eqref{eq:Z_main_app} for $n$ composite.

As an example of this general discussion, consider the case $n=12 = 2^2  \times 3$, i.e. $p_1 = 2$, $a_1 = 2$, $p_2 = 3$, and $a_2 = 1$. 12 has six divisors, i.e. 1, 2, 3, 4, 6, and 12. For each divisor $m$, the number of elements in $H_1(\Sigma, \Z_m)$ that are not contained in a smaller subgroup is (cf. Equation \eqref{eq:valueofC}) 
\bea \nonumber
\CC_{(0,0)} &=& 1 \\ \nonumber
\CC_{(1,0)} &=& 2^{2g}-1 \\ \nonumber
\CC_{(0,1)} &=& 3^{2g}-1 \\ \nonumber
\CC_{(2.0)} &=& 4^{2g} - 2^{2g} \\ \nonumber
\CC_{(1,1)} &=& 6^{2g}- 3^{2g} - 2^{2g} + 1 \\ \nonumber
\CC_{(2,1)} &=& 12^{2g} - 6^{2g} - 4^{2g} + 2^{2g}
\eea
Plugging these numbers in \eqref{eq:partition_function_2}, we obtain that the coefficient of $1$ is
\bea 
\sum\limits_{i \in B} N_i^{2g-2} - \sum\limits_{i \in \CB^{\Z_2}} N_i^{2g-2} - \sum\limits_{i \in \CB^{\Z_3}} N_i^{2g-2} + \sum\limits_{i \in \CB^{\Z_6}} N_i^{2g-2}. 
\eea
which is precisely the sum of $N_i^{2g-2}$ over the set of elements in orbits of length 12. Similarly, the coefficient of $2^{2g}$ is 
\bea 
\sum\limits_{i \in \CB^{\Z_2}} N_i^{2g-2} - \sum\limits_{i \in \CB^{\Z_4}} N_i^{2g-2} - \sum\limits_{i \in \CB^{\Z_6}} N_i^{2g-2} + \sum\limits_{i \in \CB^{\Z_{12}}} N_i^{2g-2}.
\eea
which is the sum over elements in orbits of length 6. After making similar computations for coefficients of $3^{2g}$, $4^{2g}$, $6^{2g}$ and $12^{2g}$ it is straightforward to derive that the partition function of $\CT/\Z_{12}^\zero$ is 
\bea 
\CZ_{\CT/\Z_{12}^\zero} = \sum\limits_{i \in \CB/\Z_{12}} |\Stab (i)| \left(\frac{|\Stab (i)|}{|\Z_{12}|^\half}  N_i \right)^{2g-2}
\eea

\subsection{The case of product of cyclic groups} \label{sec:product}
Finally, we will discuss the case of gauging (a subgroup of) the 0-form symmetry of a TQFT that is a product of cyclic groups. Here, we may gauge one factor of the product at a time. At each step, the partition function can be computed using the techniques presented in the previous two subsections. A special case of interest that has appeared implicitly in this paper is that of $\CT(Spin(4n,\C))$. The center of $Spin(4n,\C)$ is $\Z_2 \times \Z_2$. As this subgroup of the 0-form symmetry is gauged, a natural generalization of the results \eqref{eq:Z_h_app} and \eqref{eq:Z_main_app} holds. 

In general, for a TQFT $\CT$ with 0-form symmetry $\Z_2 \times \Z_2$, $\CZ_h$ for $h \in H_1(\Sigma, \Z_2 \times \Z_2)$ is given by
\bea \label{eq:A.14}
\CZ_h = \sum\limits_{\lambda \in P_k^{\widetilde{Z}}} (\theta_t (f_{\lambda,t}))^{1-g}
\eea
Here $\widetilde{Z} \subset \Z_2 \times \Z_2$ is the smallest subgroup such that $H_1(\Sigma, \widetilde{Z})$ contains $h$. Furthermore, 
\bea \label{eq:A.15}
\CZ_{\CT/(\Z_2 \times \Z_2)^\zero} &=& \sum\limits_{i \in \CB/(\Z_2 \times \Z_2)}|\Stab(i)| \left( \frac{|\Stab(i)|}{|\Z_2 \times \Z_2|^\half}  N_i \right)^{2g-2}. 
\eea
This result follows easily by gauging the two $\Z_2$ factors one after another, and using the result \eqref{eq:Z_main_app}. The consistency of \eqref{eq:A.14} with \eqref{eq:A.15} follows from the equation 
\bea 
\CZ_{\CT/(\Z_2 \times \Z_2)^\zero} = \frac{1}{|H_1(\Sigma, \Z_2)|^\half} \sum\limits_h \CZ_h.
\eea

\section{Fusion rules of $\mathcal{T}(PSL(2,\mathbb{C})_k)$} \label{sec:derivation}

In this appendix we will derive the fusion rules of $\CT(PSL(2,\C)_k)$ given in Section \ref{sec:PSL2Calgebra}. In \autoref{sec:Z21derivation} the fusion rules of $\CT(SL(2,\C)_k)/\Z_2^\one$ are derived, while in \autoref{sec:Z20derivation} a derivation of the fusion rules of $\CT(PSL(2,\C)_k) = \CT(SL(2,\C)_k)/(\Z_2^\zero \times \Z_2^\one)$ is given.

\subsection{Fusion rules of $\mathcal{T}(SL(2,\C)_k)/\Z_2^{(1)}$} \label{sec:Z21derivation}

The operators of $\CT(SL(2,\C)_k)/\Z_2^\one$ are
\bea 
x^j = \frac{1}{2}(w^j + w^{k-j})
\eea
for $j= 0, 1, \cdots, \frac{k}{2}$. The fusion rules are derived as follows.
\bea \nonumber \label{eq:derivation_fusion_1}
x^a \times x^b &=& \frac{1}{4} (w^a + w^{k-a}) \times (w^b + w^{k-b}) \\ \nonumber
&=& \frac{1}{4} \sum\limits_{c=0}^k \left( f^{ab}{}_c w^c + f^{a(k-b)}{}_c w^c + f^{(k-a)b}{}_c + f^{(k-a)(k-b)}{}_c w^c \right) \\ \nonumber
&=& \frac{1}{4} \sum\limits_{c=0}^k \left( f^{ab}{}_c w^c + f^{a(k-b)}{}_{k-c} w^{k-c} + f^{(k-a)b}{}_c + f^{(k-a)(k-b)}{}_{k-c} w^{k-c}  \right) \\ \nonumber
&=& \frac{1}{4} \sum\limits_{c=0}^k (f^{ab}{}_c + f^{(k-a)b}{}_c) (w^c + w^{k-c}) \\ \label{eq:an_important_equation}
&=& \sum\limits_{c=0}^{\frac{k}{2}-1} (f^{ab}{}_c + f^{(k-a)b}{}_c) x^c + f^{ab}{}_{\frac{k}{2}} x^{\frac{k}{2}} 
\eea
Here in the fourth equality, we used an identity $f^{ab}{}_{c} = f^{a(k-b)}{}_{k-c}$ which follows from the action of $\Z_2^\one$ symmetry.\footnote{Let $\psi$ be the generator of 1-form symmetry, i.e. $\psi \times \psi = \mathbb{1}$ and $\psi \times w^a = w^{k-a}$. Then two different ways of degenerating the five-punctured sphere $f^{\psi \psi a b}{}_c$ gives the required identity $f^{ab}{}_c = f^{a(k-b)}{}_{k-c}$. } 
It follows from \eqref{eq:derivation_fusion_1} that the fusion coefficients of $\mathcal{T}(SL(2,\C)_k)/\Z_2^{(1)}$ are
\bea 
(f')^{ab}{}_c &=& f^{ab}{}_c + f^{(k-a)b}{}_c \\
(f')^{ab}{}_\mu &=& f^{ab}{}_{\frac{k}{2}}.
\eea

\subsection{Further gauging the $\mathbb{Z}_2^{(0)}$ symmetry} \label{sec:Z20derivation}
Now we consider gauging the $\Z_2^\zero$ symmetry of $\mathcal{T}(SL(2,\C)_k)/\Z_2^{(1)}$. The operators of this theory are
\bea \label{eq:list}
x^0 , x^2, \cdots, x^{\frac{k}{2}-2}, x^{{\frac{k}{2}}^\one}, x^{{\frac{k}{2}}^{(2)}}
\eea
where $x^{{\frac{k}{2}}^\one} + x^{{\frac{k}{2}}^{(2)}}$ equals the $x^{\frac{k}{2}} $ operator of $\mathcal{T}(SL(2,\C)_k)/\Z_2^{(1)}$. The fusion rules are of six different kinds:
\bea \label{eq:psl2cf}
\widetilde{f}^{ab}{}_c, \quad \widetilde{f}^{ab}{}_\mu, \quad \widetilde{f}^{a\mu}{}_b, \quad \widetilde{f}^{a\mu}{}_\nu, \quad \widetilde{f}^{\mu \nu}{}_a, \quad \widetilde{f}^{\mu \nu}{}_\rho.
\eea
where $a,b,c \in \{0, 2, \cdots, \frac{k}{2}-2\}$, and $\mu,\nu, \rho \in \{\frac{k}{2}^\one$ and $\frac{k}{2}^{\two}\}$. Rewriting \eqref{eq:an_important_equation} as (when $a$ and $b$ are even, the first sum gets contribution only from even values of $c$)
\bea 
x^a \times x^b &=& \sum\limits_{\substack{c=0 \\ c \ \text{even}}}^{\frac{k}{2}-2} (f^{ab}{}_c + f^{(k-a)b}{}_c) x^c + f^{ab}{}_{\frac{k}{2}} (x^{\frac{k}{2}^\one} + x^{\frac{k}{2}^\two})
\eea
we obtain 
\bea 
\widetilde{f}^{ab}{}_c &=& f^{ab}{}_c + f^{(k-a)b}{}_c \\
\widetilde{f}^{ab}{}_\mu &=& f^{ab}{}_{\frac{k}{2}}
\eea

The next two fusion coefficients in \eqref{eq:list} encode the fusion of $x^a$ with $x^\mu$, but since $\sum\limits_\mu x^\mu = x^{\frac{k}{2}}$, it is useful to sum over $\mu$ in order to find relations with the fusion rules of $\CT(SL(2,\C)_k)$:
\bea \nonumber
\sum\limits_{\mu} (x^a \times x^\mu) = \frac{1}{2} (w^a + w^{k-a}) \times w^{k/2} &=& \frac{1}{2} \sum\limits_{\substack{b=0 \\ b \ even}}^k \left(f^{a\frac{k}{2}}{}_b w^b + f^{(k-a)\frac{k}{2}}{}_{k-b} w^{k-b} \right) \\ \nonumber
&=& \frac{1}{2} \sum\limits_{\substack{b=0 \\ b \ even}}^k f^{a\frac{k}{2}}{}_b (w^b + w^{k-b})  \\ \nonumber 
&=& \sum\limits_{\substack{b=0 \\ b \ even}}^{\frac{k}{2}-2}  f^{a\frac{k}{2}}{}_b  (w^b + w^{k-b}) + f^{a\frac{k}{2}}{}_{\frac{k}{2}} w^{\frac{k}{2}} \\ 
&=& \sum\limits_{\substack{b=0 \\ b \ even}}^{\frac{k}{2}-2} 2 f^{a\frac{k}{2}}{}_b x^b + f^{a\frac{k}{2}}{}_{\frac{k}{2}} (x^{\frac{k}{2}} + x^{\widetilde{\frac{k}{2}}})
\eea
It follows that $
\sum\limits_{\mu} \widetilde{f}^{a\mu}{}_b = 2 f^{a\frac{k}{2}}{}_b $ and 
$
\sum\limits_{\mu} \widetilde{f}^{a\mu}{}_{{\frac{k}{2}}^\one} =\sum\limits_{\mu} \widetilde{f}^{a\mu}{}_{{\frac{k}{2}}^\two}= f^{a\frac{k}{2}}{}_{\frac{k}{2}}$. In the former case, the dual $\Z_2^{(0)}$ symmetry of $\CT(PSL(2,\C)_k)$ gives another constraint, i.e. 
$\widetilde{f}^{a\frac{k}{2}^\one}{}_b = \widetilde{f}^{a\frac{k}{2}^\two}{}_b$, and therefore, 
\bea 
\widetilde{f}^{a\mu}{}_b = f^{a\frac{k}{2}}{}_b.
\eea
In the latter case, the dual $\Z_2^\zero$ symmetry gives the constraints $\widetilde{f}^{a\frac{k}{2}^\one}{}_{\frac{k}{2}^\one} = \widetilde{f}^{a\frac{k}{2}^\two}{}_{\frac{k}{2}^\two}$, and $\widetilde{f}^{a\frac{k}{2}^\one}{}_{\frac{k}{2}^\two} = \widetilde{f}^{a\frac{k}{2}^\two}{}_{\frac{k}{2}^\one}$. However, these constraints do not fully determine $\widetilde{f}^{a\mu}{}_\nu$ as we have only three relations between four variables. In order to find these fusion coefficients uniquely, we can use the associativity and positivity conditions mentioned in Section \ref{sec:PSL2Calgebra}. The final result is given there. 

Finally, the last two fusion coefficients of \eqref{eq:list} encode the fusion of $x^\mu$ with $x^\nu$. Summing over $\mu$ and $\nu$ gives
\bea \nonumber
\sum\limits_{\mu, \nu = \frac{k}{2}, \widetilde{\frac{k}{2}} } x^\mu \times x^\nu = w^{k/2} \times w^{k/2} 
= \sum\limits_{\substack{c=0 \\ c\ even}}^{k} f^{\frac{k}{2} \frac{k}{2}}{}_c w^c 
= \sum\limits_{\substack{c=0 \\ c \ even}}^{\frac{k}{2}-2} 2 f^{\frac{k}{2}\frac{k}{2}}{}_c x^c + f^{\frac{k}{2}\frac{k}{2}}{}_{\frac{k}{2}} (x^{\frac{k}{2}} + x^{\widetilde{\frac{k}{2}}})
\eea
which is equivalent to the following conditions on $\widetilde{f}^{\mu \nu}{}_c$ and $\widetilde{f}^{\mu \nu}{}_{\rho}$:
\bea 
\sum\limits_{\mu, \nu} \widetilde{f}^{\mu \nu}{}_c &=& 2f^{\frac{k}{2} \frac{k}{2}}{}_c \\
\sum\limits_{\mu, \nu} \widetilde{f}^{\mu \nu}{}_{\frac{k}{2}} &=& \sum\limits_{\mu, \nu} \widetilde{f}^{\mu \nu}{}_{\widetilde{\frac{k}{2}}} = f^{\frac{k}{2}\frac{k}{2}}{}_{\frac{k}{2}} 
\eea
Again, we can obtain more constraints by using the dual $\Z_2^\zero$ symmetry of $\CT(PSL(2,\C)_k)$, i.e. 
\bea \nonumber
\widetilde{f}^{\frac{k}{2}^\one \frac{k}{2}^\one}{}_{a} =  
\widetilde{f}^{\frac{k}{2}^\two \frac{k}{2}^\two}{}_{a} \quad ; \quad 
\widetilde{f}^{\frac{k}{2}^\one \frac{k}{2}^\two}{}_{a} = 
\widetilde{f}^{\frac{k}{2}^\two \frac{k}{2}^\one}{}_{a} 
\eea
and 
\bea 
\widetilde{f}^{\frac{k}{2}^\one \frac{k}{2}^\one}{}_{\frac{k}{2}^\one} = 
\widetilde{f}^{\frac{k}{2}^\two \frac{k}{2}^\two}{}_{\frac{k}{2}^\two}  \quad ; \quad
\widetilde{f}^{\frac{k}{2}^\one \frac{k}{2}^\two}{}_{\frac{k}{2}^\one} = 
\widetilde{f}^{\frac{k}{2}^\two \frac{k}{2}^\one}{}_{\frac{k}{2}^\two}
\eea
However, just like in the case of $\widetilde{f}^{a\mu}{}_\nu$, these constraints are not enough to fix the fusion coefficients entirely. Using associativity and positivity constraints, the results can be found and were given in Section \ref{sec:PSL2Calgebra}. 

\bibliographystyle{utphys}
\bibliography{biblio}

\providecommand{\href}[2]{#2}\begingroup\raggedright\begin{thebibliography}{10}

\bibitem{VERLINDE1988360}
E.~Verlinde, ``{Fusion rules and modular transformations in 2D conformal field
  theory},'' {\em Nuclear Physics B} {\bf 300} (1988) 360--376.

\bibitem{Witten:1993xi}
E.~Witten, ``{The Verlinde algebra and the cohomology of the Grassmannian},''
  \href{http://www.arXiv.org/abs/hep-th/9312104}{{\tt hep-th/9312104}}.

\bibitem{cmp/1104248305}
D.~Gepner, ``{Fusion rings and geometry},'' {\em Communications in Mathematical
  Physics} {\bf 141} (1991), no.~2, 381 -- 411.

\bibitem{cmp/1104248198}
E.~Witten, ``{{On quantum gauge theories in two dimensions}},'' {\em
  Communications in Mathematical Physics} {\bf 141} (1991), no.~1, 153 -- 209.

\bibitem{Witten:1992xu}
E.~Witten, ``{Two-dimensional gauge theories revisited},'' {\em J. Geom. Phys.}
  {\bf 9} (1992) 303--368, \href{http://www.arXiv.org/abs/hep-th/9204083}{{\tt
  hep-th/9204083}}.

\bibitem{Gukov:2015sna}
S.~Gukov and D.~Pei, ``{Equivariant Verlinde formula from fivebranes and
  vortices},'' {\em Commun. Math. Phys.} {\bf 355} (2017), no.~1, 1--50,
  \href{http://www.arXiv.org/abs/1501.01310}{{\tt 1501.01310}}.

\bibitem{Andersen:2016hoj}
J.~E. Andersen, S.~Gukov, and D.~Pei, ``{The Verlinde formula for Higgs
  bundles},'' \href{http://www.arXiv.org/abs/1608.01761}{{\tt 1608.01761}}.

\bibitem{halpernleistner2016equivariant}
D.~Halpern-Leistner, ``{The equivariant Verlinde formula on the moduli of Higgs
  bundles},'' 2016.

\bibitem{Gaiotto:2014kfa}
D.~Gaiotto, A.~Kapustin, N.~Seiberg, and B.~Willett, ``{Generalized Global
  Symmetries},'' {\em JHEP} {\bf 02} (2015) 172,
  \href{http://www.arXiv.org/abs/1412.5148}{{\tt 1412.5148}}.

\bibitem{Durhuus:1993cq}
B.~Durhuus and T.~Jonsson, ``{Classification and construction of unitary
  topological field theories in two-dimensions},'' {\em J. Math. Phys.} {\bf
  35} (1994) 5306--5313, \href{http://www.arXiv.org/abs/hep-th/9308043}{{\tt
  hep-th/9308043}}.

\bibitem{Gukov:2016lki}
S.~Gukov, D.~Pei, W.~Yan, and K.~Ye, ``{Equivariant Verlinde Algebra from
  Superconformal Index and Argyres\textendash{}Seiberg Duality},'' {\em Commun.
  Math. Phys.} {\bf 357} (2018), no.~3, 1215--1251,
  \href{http://www.arXiv.org/abs/1605.06528}{{\tt 1605.06528}}.

\bibitem{10.1215/S0012-7094-94-07618-7}
T.~Pantev, ``{{Comparison of generalized theta functions}},'' {\em Duke
  Mathematical Journal} {\bf 76} (1994), no.~2, 509 -- 539.

\bibitem{beauville1996verlinde}
A.~Beauville, ``{The Verlinde formula for PGL (p)},'' {\em arXiv preprint
  alg-geom/9609017} (1996).

\bibitem{meinrenken2018verlinde}
E.~Meinrenken, ``{Verlinde formulas for nonsimply connected groups},'' in {\em
  Lie Groups, Geometry, and Representation Theory}, pp.~381--417.
\newblock Springer, 2018.

\bibitem{10.2307/1970717}
M.~F. Atiyah and I.~M. Singer, ``{The Index of Elliptic Operators: III},'' {\em
  Annals of Mathematics} {\bf 87} (1968), no.~3, 546--604.

\bibitem{Cordova:2019jnf}
C.~C\'ordova, D.~S. Freed, H.~T. Lam, and N.~Seiberg, ``{Anomalies in the Space
  of Coupling Constants and Their Dynamical Applications I},'' {\em SciPost
  Phys.} {\bf 8} (2020), no.~1, 001,
  \href{http://www.arXiv.org/abs/1905.09315}{{\tt 1905.09315}}.

\bibitem{Cordova:2019uob}
C.~C\'ordova, D.~S. Freed, H.~T. Lam, and N.~Seiberg, ``{Anomalies in the Space
  of Coupling Constants and Their Dynamical Applications II},'' {\em SciPost
  Phys.} {\bf 8} (2020), no.~1, 002,
  \href{http://www.arXiv.org/abs/1905.13361}{{\tt 1905.13361}}.

\bibitem{Sharpe:2000ki}
E.~R. Sharpe, ``{Discrete torsion},'' {\em Phys. Rev. D} {\bf 68} (2003)
  126003, \href{http://www.arXiv.org/abs/hep-th/0008154}{{\tt hep-th/0008154}}.

\bibitem{Hellerman:2006zs}
S.~Hellerman, A.~Henriques, T.~Pantev, E.~Sharpe, and M.~Ando, ``{Cluster
  decomposition, T-duality, and gerby CFT's},'' {\em Adv. Theor. Math. Phys.}
  {\bf 11} (2007), no.~5, 751--818,
  \href{http://www.arXiv.org/abs/hep-th/0606034}{{\tt hep-th/0606034}}.

\bibitem{Moore:2006dw}
G.~W. Moore and G.~Segal, ``{D-branes and K-theory in 2D topological field
  theory},'' \href{http://www.arXiv.org/abs/hep-th/0609042}{{\tt
  hep-th/0609042}}.

\bibitem{teleman2009index}
C.~Teleman and C.~T. Woodward, ``The index formula for the moduli of g-bundles
  on a curve,'' {\em Annals of mathematics} (2009) 495--527.

\bibitem{beauville1994conformal}
A.~Beauville, ``{Conformal blocks, fusion rules and the Verlinde formula},''
  {\em arXiv preprint alg-geom/9405001} (1994).

\bibitem{Moore:1997dj}
G.~W. Moore, N.~Nekrasov, and S.~Shatashvili, ``{Integrating over Higgs
  branches},'' {\em Commun. Math. Phys.} {\bf 209} (2000) 97--121,
  \href{http://www.arXiv.org/abs/hep-th/9712241}{{\tt hep-th/9712241}}.

\bibitem{Nekrasov:2014xaa}
N.~A. Nekrasov and S.~L. Shatashvili, ``{Bethe/Gauge correspondence on curved
  spaces},'' {\em JHEP} {\bf 01} (2015) 100,
  \href{http://www.arXiv.org/abs/1405.6046}{{\tt 1405.6046}}.

\bibitem{Okuda:2013fea}
S.~Okuda and Y.~Yoshida, ``{G/G gauged WZW-matter model, Bethe Ansatz for
  q-boson model and Commutative Frobenius algebra},'' {\em JHEP} {\bf 03}
  (2014) 003, \href{http://www.arXiv.org/abs/1308.4608}{{\tt 1308.4608}}.

\bibitem{Benini:2016hjo}
F.~Benini and A.~Zaffaroni, ``{Supersymmetric partition functions on Riemann
  surfaces},'' {\em Proc. Symp. Pure Math.} {\bf 96} (2017) 13--46,
  \href{http://www.arXiv.org/abs/1605.06120}{{\tt 1605.06120}}.

\bibitem{Closset:2016arn}
C.~Closset and H.~Kim, ``{Comments on twisted indices in 3d supersymmetric
  gauge theories},'' {\em JHEP} {\bf 08} (2016) 059,
  \href{http://www.arXiv.org/abs/1605.06531}{{\tt 1605.06531}}.

\bibitem{Kanno:2018qbn}
H.~Kanno, K.~Sugiyama, and Y.~Yoshida, ``{Equivariant U(N) Verlinde algebra
  from Bethe/Gauge correspondence},'' {\em JHEP} {\bf 02} (2019) 097,
  \href{http://www.arXiv.org/abs/1806.03039}{{\tt 1806.03039}}.

\bibitem{Ueda:2019qhg}
K.~Ueda and Y.~Yoshida, ``{3d $ \mathcal{N} $ = 2 Chern-Simons-matter theory,
  Bethe ansatz, and quantum $K$-theory of Grassmannians},'' {\em JHEP} {\bf 08}
  (2020) 157, \href{http://www.arXiv.org/abs/1912.03792}{{\tt 1912.03792}}.

\bibitem{eckhard2020higher}
J.~Eckhard, H.~Kim, S.~Sch{\"a}fer-Nameki, and B.~Willett, ``Higher-form
  symmetries, bethe vacua, and the 3d-3d correspondence,'' {\em Journal of High
  Energy Physics} {\bf 2020} (2020), no.~1, 1--90.

\bibitem{Closset:2019hyt}
C.~Closset and H.~Kim, ``{Three-dimensional \ensuremath{\mathscr{N}} = 2
  supersymmetric gauge theories and partition functions on Seifert manifolds: A
  review},'' {\em Int. J. Mod. Phys. A} {\bf 34} (2019), no.~23, 1930011,
  \href{http://www.arXiv.org/abs/1908.08875}{{\tt 1908.08875}}.

\bibitem{Moore:1989yh}
G.~W. Moore and N.~Seiberg, ``{Taming the Conformal Zoo},'' {\em Phys. Lett. B}
  {\bf 220} (1989) 422--430.

\bibitem{Dijkgraaf:1989pz}
R.~Dijkgraaf and E.~Witten, ``{Topological Gauge Theories and Group
  Cohomology},'' {\em Commun. Math. Phys.} {\bf 129} (1990) 393.

\bibitem{laredo1999positive}
V.~T. Laredo, ``{Positive energy representations of the loop groups of
  non-simply connected Lie groups},'' {\em Communications in mathematical
  physics} {\bf 207} (1999), no.~2, 307--339.

\bibitem{DiFrancesco:1997nk}
P.~Di~Francesco, P.~Mathieu, and D.~Senechal, {\em {Conformal Field Theory}}.
\newblock Graduate Texts in Contemporary Physics. Springer-Verlag, New York,
  1997.

\bibitem{Delmastro:2021xox}
D.~Delmastro, D.~Gaiotto, and J.~Gomis, ``{Global Anomalies on the Hilbert
  Space},'' \href{http://www.arXiv.org/abs/2101.02218}{{\tt 2101.02218}}.

\bibitem{Brustein:1988vb}
R.~Brustein, S.~Yankielowicz, and J.-B. Zuber, ``{Factorization and Selection
  Rules of Operator Product Algebras in Conformal Field Theories},'' {\em Nucl.
  Phys. B} {\bf 313} (1989) 321--347.

\bibitem{RISR2011gerbe}
I.~Runkel and R.~R. Suszek, ``{Affine su(2) fusion rules from gerbe
  2-isomorphisms},'' {\em Journal of Geometry and Physics} {\bf 61} (Aug, 2011)
  1527–1552.

\bibitem{Gukov:2010sw}
S.~Gukov, ``{Quantization via Mirror Symmetry},''
  \href{http://www.arXiv.org/abs/1011.2218}{{\tt 1011.2218}}.

\end{thebibliography}\endgroup

\end{document}